\documentstyle[aps,floats,epsf,twocolumn,eqsecnum,amsmath,bm]{revtex}
\newcommand{\bk}{{\bf k}}

\newcommand{\br}{{\bf r}}
\newcommand{\bR}{{\bf R}}
\newcommand{\bx}{{\bf x}}
\newcommand{\by}{{\bf y}}

\newcommand{\bA}{{\bf A}}
\newcommand{\bv}{{\bf v}}
\newcommand{\bB}{{\bf B}}
\newcommand{\bp}{{\bf p}}

\newcommand{\bl}{{\bf l}}

\newcommand{\bdelta}{\bm{\delta}}
\newcommand{\bpi}{\bm{\pi}}
\newcommand{\eps}{\epsilon}
\newcommand{\pibar}{{\pi}^*}
\newcommand{\bpibar}{{\bpi}^*}

\newcommand{\bj}{{\bf j}}

\newcommand{\psidag}{\psi^{\dagger}}
\newcommand{\Psidag}{\Psi^{\dagger}}
\newcommand{\Psidot}{\dot{\Psi}}
\newcommand{\Psidotdag}{\dot{\Psi}^{\dagger}}
\newcommand{\Psitilda}{\tilde{\Psi}}
\newcommand{\Psitildadag}{\tilde{\Psi}^{\dagger}}
\newcommand{\Psitildadot}{\dot{\tilde{\Psi}}}
\newcommand{\Psitildadotdag}{\dot{\tilde{\Psi}}^{\dagger}}

\newcommand{\Deltaop}{{\hat{\Delta}}}

\newcommand{\calP}{{\cal P}}

\newcommand{\hatG}{\hat{G}}
\newcommand{\hatA}{\hat{A}}
\newcommand{\hH}{\hat{H}_0}

\newcommand{\chifac}{{(1+\frac\chi{2})}}
\newcommand{\bra}{\langle}
\newcommand{\ket}{\rangle}
\newcommand{\bigbra}{\biggl{\langle}}
\newcommand{\bigket}{\biggr{\rangle}}
\newcommand{\Deltaab}{{\Delta_{\alpha\beta}}}
\newcommand{\Tr}{{\rm Tr }}
\newcommand{\tL}{\tilde{L}}

\begin{document} \draft

\title{Quasiparticle Hall Transport of d-wave Superconductors in Vortex State}

\author{O. Vafek, A. Melikyan, and Z. Te\v{s}anovi\'c}
\address{Department of Physics and Astronomy, Johns Hopkins University,
Baltimore, MD 21218
\\ {\rm(\today)}
}
\address{~
\parbox{14cm}{\rm
\medskip
We present a theory of quasiparticle Hall
transport in strongly type-II superconductors
within their vortex state.
We establish the existence of integer quantum spin Hall effect in clean
unconventional $d_{x^2-y^2}$ superconductors in the vortex state
from a general analysis of the Bogoliubov-de~Gennes equation.
The spin Hall conductivity $\sigma^s_{xy}$ is shown
to be quantized in units of $\frac{\hbar}{8\pi}$.
This result does not rest on linearization of the
BdG equations around Dirac nodes and therefore
includes inter-nodal physics in its entirety. In addition,
this result holds for a generic inversion-symmetric
lattice of vortices as long as the magnetic field $B$
satisfies $H_{c1} \ll B \ll H_{c2}$.
We then derive the Wiedemann-Franz law for the spin and
thermal Hall conductivity in the vortex state.
In the limit of $T \rightarrow 0$, the thermal Hall conductivity satisfies
$\kappa_{x y}=\frac{4\pi^2}{3}(\frac{ k_B}{\hbar})^2 T \sigma^s_{xy}$.
The transitions between different quantized
values of $\sigma^s_{xy}$ as well as relation to
conventional superconductors are discussed.
}}

\maketitle

\pacs{74.60.-w,74.60.Ec,74.72.-h}
\section{Introduction}
One of the fundamental characteristics of high temperature
superconductors (HTS) is the apparent
applicability of the d-wave \cite{harlingen} BCS based phenomenology
to the broad range of quasiparticle
properties in the superconducting state\cite{millis}.
This is far from trivial property for materials known to
exhibit strong electron correlations.
Although the horizon is still not entirely clear and there
remain few unresolved issues,
examples being the temperature dependence of quasiparticle lifetimes
or the penetration depth in the under-doped regime,\cite{millis,walker}
the cumulative weight of evidence indicates
that the low energy properties of cuprate
superconductors are indeed governed
by nodal quasiparticles with Dirac like dispersion, as seen in
assorted spectroscopic \cite{ding} and transport
measurements \cite{taillefert1}.
These and other experiments serve as the foundation for
the ``BCS-like d-wave paradigm'' as it is currently used in
both theory and interpretation of experiments.

A natural question is how does this picture hold together in a mixed
phase, in presence of an external magnetic field
and an array of superconducting vortices and, if it
does, are there some special features of the d-wave quasiparticle
phenomenology which could be used to deepen our understanding
of high temperature superconductivity.
Recent activity on the experimental front appears most
encouraging in this regard.
In particular,  measurements of the thermal Hall
conductivity $\kappa_{xy}$ in cuprate superconductors are currently
viewed as especially informative probes of quasiparticle
dynamics. These measurements provide a clean way of extracting
quasiparticle contribution to $\kappa_{xy}$ as phonons, the
other source of significant thermal conduction,
do not couple to the magnetic field by virtue of being neutral.
In addition, contrary to what takes place in an
ordinary electrical Hall conductivity experiment,
vortices do not experience strong Lorenz force since
there is only  heat current and no net electrical supercurrent
to which vortices are strongly coupled.
Thus, vortices tend to remain stationary and their transport
does not serve as a significant channel for heat conduction.

Recently, measurements of $\kappa_{xy}$ were conducted by Ong's group \cite{ong}
on YBCO samples with a very long mean free path.
These experiments were carried out
over a wide range of magnetic fields
(up to $\sim 14T$) and at temperatures from $T \sim 12.5K$ to above the
superconducting transition
$T_c \sim 90K$. Unfortunately, the experiments
are resolution limited below
$12.5K$ as signal becomes too weak. At temperatures up
to $25K$ and for a (T dependent)
range of magnetic fields $\sqrt{H/Tesla} < T/25K$ the
experiments seem to suggest a
rather simple scaling form \cite{simonlee} for $\kappa_{xy}$:
\begin{equation}
\kappa_{xy}(H,T) = const.\times  \sqrt{H} \;T.
\end{equation}

On the theoretical front, the initial interest, largely inspired by
Gorkov and Schrieffer \cite{gorkov} and, in a somewhat different context,
by Anderson \cite{anderson}, was directed
at the formation of ``Dirac Landau levels" and their signatures in the
quasiparticle thermodynamics and transport \cite{gusynin,kopnin}.
This picture of Dirac Landau level quantization, while theoretically
elegant and appealing, relies on the minimal
coupling of the nodal BCS-like
quasiparticles to the electromagnetic vector potential, ${\bf A}$.
The assumption of such minimal coupling
seems innocent but it is not, on fundamental grounds.
The physics behind the interaction of nodal quasiparticles with
the external magnetic field and vortices was elucidated by
Franz and  Te\v{s}anovi\'c \cite{ft}. These authors devised a
singular gauge transformation which allows one to recast the original
problem of BCS-like quasiparticles, moving under the combined
influence of an external magnetic field and a superflow arising from
vortex array, into that of Bloch particles moving in an effective
non-uniform and periodic magnetic field, the
spatial average of which equals zero.
This approach clearly demonstrates that the low energy portion of the
quasiparticle spectrum can be
described as that of a relativistic Dirac particle minimally
coupled to a fictitious U(1) gauge potential, i.e. the
``Berry'' gauge field {\bf a}, which supplies the needed $\pm\pi$
winding in the quantum mechanical phase of a quasiparticle as
it encircles a vortex. Such half-flux Aharonov-Bohm scattering
arises entirely through interaction of quasiparticles with
vortices and it does not involve the external magnetic field
explicitly.
Thus, the cyclotron motion in a Dirac cone is caused exclusively by
a {\em time-reversal invariant} ``Berry'' gauge field and
cannot lead to any Dirac Landau level quantization.
Further progress came through the work of
Marinelli, Halperin and Simon\cite{marinelli}
who analyzed the quasiparticle excitation spectra of different
vortex lattices and provided analytic symmetry arguments
regarding the presence of nodal (zero energy) points in
such spectra. These authors also devised a perturbation
theory in the vicinity of nodal points which can be used
to derive various results by analytic means. Various analytic
results were also derived in the large anisotropy
limit \cite{Melnikov,kallin}.
Finally, the intricacies of Dirac equation in presence of half-flux
Bohm-Aharonov scattering were addressed in Ref. \cite{vmft}, where
the tight-binding regularization was introduced to restore
the exact singular gauge symmetry in numerical calculations
using the low energy (linearized) theory. Such regularization
describes a lattice d-wave superconductor and
is a natural choice for cuprates:  after all, most of the
microscopic theories of HTS start from tight-binding effective
Hamiltonians. We should note that similar
results for the quasiparticle spectra were also found in fully
self-consistent calculations on the original
Bogoliubov-deGennes (BdG) equations, using the basis formed by
eigenstates of the magnetic translation group \cite{wang1,kita1}.

In this paper, we present a detailed analytical study of the
quasiparticle Hall transport in a vortex state complemented by
explicit numerical calculations \cite{ashvin}.
We consider a lattice d-wave superconductor of
Ref. \cite{vmft} -- This is important and necessary since
the straightforward linearization of BdG equations
drops curvature terms and results in $\kappa_{xy} = 0$\cite{simonlee,ye}.
We employ the Franz and  Te\v{s}anovi\'c (FT) transformation
so that we can use the familiar Bloch representation of
the translation group in which the overall chirality of the
problem vanishes. This should be contrasted
with the original problem where
the overall chirality is finite and the magnetic translation
group states must be used instead.
Naively, it might appear that after an FT
singular gauge transformation the
effects of the magnetic field have somehow been
transformed away since the new problem
is found to have zero average effective
magnetic field. Of course, this is not true.
The presence of magnetic field in the original problem
reveals itself fully in the FT transformed
quasiparticle wavefunctions.
Alternatively, there is an ``intrinsic'' chirality
imposed on the system which cannot
be transformed away by a choice of the basis.
One manifestation of this chirality is
the Hall effect. The utility of the singular
gauge transformation in the calculation of the electrical Hall conductivity
in the {\em normal} 2D electron gas in
a (non-uniform) magnetic field was realized
by Nielsen and Hedeg{\aa}rd \cite{nielsen}.
They demonstrated that using singularly
gauge transformed wavefunctions one still obtains the correct
result, giving the electrical Hall conductance
quantized in units of $e^2/h$ if the chemical potential lies in the
energy gap.
In a superconductor, the question of Hall response becomes
rather interesting  as
there is a strong mixing between particles and holes.
Evidently, the electrical
Hall response is very different from the normal state, since charge is not
conserved in the state with broken U(1) symmetry.
Therefore, as pointed out in Ref. \cite{senthil},
charge cannot be transported by diffusion.
On the other hand, the spin is still a good quantum number\cite{senthil}
and it is natural to ask
what is the spin Hall conductivity in the vortex state of an extreme type-II
superconductor \cite{volovik}.
Moreover, every channel of spin conduction
simultaneously transports
entropy \cite{senthil,durst,ashvin} and we would expect
some variation on Wiedemann-Franz law to hold between
spin and thermal conductivity.

As one of our main results, we derive the Wiedemann-Franz law connecting
the spin and thermal Hall transport in the {\em vortex state}
of a d-wave superconductor. In the process, we show that the
spin Hall conductivity,  $\sigma_{xy}^s$,
just like the electrical Hall conductivity of
a normal state in a magnetic field, is topological in nature
and can be explicitly evaluated as a first Chern number
characterizing the eigenstates of our singularly gauge transformed
problem \cite{haldane,volovik,read1}. Consequently, as
$T \rightarrow 0$, the spin Hall conductivity
is quantized in the units of $\hbar/8\pi$
when the energy spectrum is gapped,
which, combined with the Wiedemann-Franz law,
implies the quantization of $\kappa_{xy}/T$. We then explicitly
compute the quantized values of $\sigma_{xy}^s$ for a sequence
of gapped states using our lattice d-wave superconductor model
in the case of an inversion-symmetric vortex lattice.
Within this model one is naturally led to consider the
{\em BCS-Hofstadter} problem: the BCS pairing problem defined on
a uniformly frustrated tight-binding lattice. We find
a sequence of plateau transitions, separating gapped states
characterized by different quantized values of $\sigma_{xy}^s$.
At a plateau transition, level crossings take place
and $\sigma_{xy}^s$ changes by an even integer \cite{even}. 
Both the origin of the gaps in quasiparticle
spectra and the sequence of values for $\sigma_{xy}^s$
are rather different than in the normal state, i.e. in
the standard Hofstadter problem \cite{morita}. In a superconductor,
the gaps are strongly affected by the pairing and the
interactions of quasiparticles with a vortex array.
The sequence of $\sigma_{xy}^s$ changes as a function
of the pairing strength (and therefore interactions),
measured by the maximum value
of the gap function $\Delta$ \cite{read2}. Finally,
we discuss the relation of our results to those
obtained within the continuum low energy (linearized) theory.

\section{Bogoliubov-deGennes Hamiltonian}
The experimental evidence points towards well defined d-wave quasiparticles in cuprate superconductors in the absence
of the external magnetic field. This suggests that to zeroth order fluctuations
can be ignored and that one can think in terms of an effective BCS Hamiltonian, the simplest of which is written on
the 2-D tight-binding lattice with the nearest neighbor interaction thus naturally
implementing $d_{x^2-y^2}$ pairing.
In question is then the response of such a superconductor to an externally applied magnetic field $\bB$.
All high temperature superconductors are extreme type-II forming a vortex state in a wide range of
magnetic fields. This immediately sets up the contrast between $\bB=0$ and $\bB \neq 0$ situations:
first, the problem is not spatially uniform and therefore momentum is not a
good quantum number and second, the array of $\frac{hc}{2e}$ vortex fluxes poses topological
constraint on the quasi-particles encircling the vortices. Therefore, despite ignoring any
fluctuations, the problem is far from trivial and demands careful examination.

The natural starting point is therefore the mean-field BCS Hamiltonian written in second quantized form \cite{deGennes89}:
\begin{multline}
H=\int d {\bf x} \;\psi^{\dagger}_{\alpha}(\bx)
\biggr{(}\frac{1}{2m^{*}}(\bp-\frac{e}{c}{\bf A})^2 -\mu \biggl{)}\psi_{\alpha}(\bx)+\\
\int d {\bf x} \int d {\bf y}[ \Delta({\bf x},{\bf y})
\psi_{\uparrow}^{\dagger}(\bx)
\psi_{\downarrow}^{\dagger}(\by)+\Delta^{*}({\bf x},{\bf y})
\psi_{\downarrow}(\by) \psi_{\uparrow}(\bx)   ]
\label{bcshamiltonian}
\end{multline}
where ${\bf A}({\bf x})$ is the vector
potential associated with the uniform external magnetic field $\bB$,
single electron energy is measured relative to the chemical potential $\mu$,
$\psi_{\alpha}(\bx)$ is the fermion field operator with spin index $\alpha$,
and $\Delta({\bf x},{\bf y})$ is the pairing field.
For convenience we will define an integral operator $\hat{\Delta}$ such that:
\begin{equation}
\hat{\Delta} \psi({\bf x})= \int d {\bf y} \Delta(\bx,\by) \psi({\bf y}).
\end{equation}
In the strictest sense, on the mean field level this problem must be solved self-consistently
which renders any analytical solution virtually intractable.
On the other hand, in the case at hand the
vortex lattice is dilute for a wide range of magnetic fields, and
by the very nature of cuprate superconductors having short coherence length,
the size of the vortex core can be 
ignored relative to the distance between the vortices.
Thus, to the first approximation, all essential physics is captured by fixing
the amplitude of the
order parameter $\Delta$ while allowing vortex defects in its phase.
Moreover, on a tight-binding lattice the vortex flux is concentrated
inside the plaquette and thus the length-scale
associated with the core is implicitly the lattice spacing $\bdelta$ of the underlying
tight-binding lattice.
As shown in Ref.\cite{vmft}, under these approximations the d-wave pairing operator
in the vortex state can be written as a differential operator:
\begin{equation}
\hat{\Delta}= \Delta_{0} \sum_{{\bf \delta}}\eta_{\bdelta}e^{i \phi({\bx})/2}\;
 e^{i \bdelta \cdot \bp}\;e^{i \phi({\bf x})/2}.
\label{deltaoperator}
\end{equation}
The sums are over nearest neighbors and on the square
tight-binding lattice $\bdelta= \pm \hat{x}, \pm \hat{y}$;
the vortex phase fields satisfy
$\nabla\times\nabla\phi ({\bx}) = 2\pi\hat{z}\sum_i
\delta ({\bx} -{\bx}_i)$ with $\bx_i$ denoting the vortex positions
and $\delta ({\bx} -{\bx}_i) $ being a 2D Dirac delta function;
$\bp$ is a momentum operator, and
\begin{equation}
 \eta_{\bdelta}= \left\{ \begin{array}{ll}
     1 & \mbox{if $ \bdelta  = \pm \hat{x}$ } \\
    -1 & \mbox{if $ \bdelta = \pm \hat{y}$}.
\end{array} \right.
\label{etaoperator}
\end{equation}
The operator $\eta_{\bdelta}$ follows from the
$d$-wave pairing:
$\Delta = 2 \Delta_0 [\cos(k_x \delta_x) -\cos(k_y \delta_y)].$
For notational convenience we will use units where $\hbar=1$ and return to
the conventional units when necessary.

It is straightforward to derive the continuum version of the tight binding
lattice operator $\hat{\Delta}$ (see Ref.\cite{vmft}):
\begin{eqnarray}
\hat{\Delta}=\frac{1}{2p_F^2}\{\partial_x,\{\partial_x,\Delta (\bx)\}\} -
\frac{1}{2p_F^2}\{\partial_y,\{\partial_y,\Delta (\bx)\}\} + \nonumber \\
+\frac{i}{8p_F^2}\Delta (\bx)\bigl [(\partial_x^2\phi) - (\partial_y^2\phi)\bigr ],
\label{avii}
\end{eqnarray}
but for convenience we will keep the
lattice definition (\ref{deltaoperator}) throughout.
One can always define continuum as an appropriate limit of the
tight-binding lattice theory.
With the above definitions, the Hamiltonian (\ref{bcshamiltonian})
can now be written in the Nambu formalism as
\begin{equation}
H=\int d {\bf x} \;\; \Psi^{\dagger}(\bx)\; \hat{H}_0 \;\Psi(\bx)
\label{bcsnambu}
\end{equation}
where the Nambu spinor $\Psi^{\dagger}=(\psi^{\dagger}_{\uparrow}, \psi_{\downarrow})$
and the matrix differential operator
\begin{equation}
\hat{H}_0
=\left( \begin{array}{cc}
\hat{h} & \hat{\Delta} \\
\hat{\Delta}^{\ast} & -\hat{h}^{\ast}
\end{array} \right).
\label{bcshamnambu}
\end{equation}
In the continuum formulation $\hat{h}=\frac{1}{2m^{*}}({\bf
p}-\frac{e}{c}{\bf A})^2-\mu$, while on the tight-binding lattice:
\begin{equation}
\hat{h} = -t \sum_{{\bf \delta}}e^{i\int_{\bx}^{{\bx}
+ {\bf \delta}}(\bp- \frac{e}{c}{\bf A})\cdot d{\bf l}}\;-
\mu.
\label{tbkenergy}
\end{equation}
$t$ is the hopping constant and $\mu$ is the Fermi energy.
The equations of motion of the Nambu fields $\Psi$ are then:
\begin{equation}
\label{eqmot}
i\hbar \dot{\Psi}=[\Psi,H] =\hat{H}_0 \Psi.
\end{equation}

Note, that the Hamiltonian in Eq. (\ref{bcshamiltonian})
is our starting {\it unperturbed} Hamiltonian.
In order to compute the linear response to externally applied perturbations
we will have to add terms to Eq. (\ref{bcshamiltonian}).
In particular, we will consider two types of perturbations in the later sections:
First, partly for theoretical convenience,
we will consider a weak gradient of magnetic field ($\nabla \bB$)
on top of the uniform $\bB$ already
taken into account fully by Eq. (\ref{bcshamiltonian}).
The $\nabla \bB$ term induces spin current
in the superconductor \cite{senthil2}. The response is
then characterized by spin conductivity tensor $\sigma^s$
which in general has non-zero off-diagonal components.
Second, we consider perturbing the system by
pseudo-gravitational field, which formally induces flow of energy
(see \cite{luttinger,obraz,smst,streda})
and allows us to compute thermal conductivity $\kappa_{xy}$ via linear response.
The advantage of these formal considerations are made clear in Section IV.
\subsection{Particle-Hole Symmetry}
The equations of motion (\ref{eqmot}) for stationary states
lead to Bogoliubov-de~Gennes equations \cite{vmft}
\begin{equation}
{\hat{H}_0}\Phi_n = \eps_n \Phi_n.
\label{bdg}
\end{equation}
The solution of these coupled differential equations are quasi-particle
wavefunctions that are rank two spinors in the Nambu space,
${\Phi}^T({\bf r})=(u({\bf r}),v({\bf r}))$. The single particle excitations
of the system are completely specified once the quasi-particle
wavefunctions are given, and as discussed later, transverse transport coefficients
can be computed solely on the basis of $\Phi$'s.
It is a general symmetry of the BdG equations
that if $(u_n(\br), v_n(\br))$ is a solution with energy ${\eps}_n$,
then there is always another solution $(-v^*_n(\br), u^*_n(\br))$ with
energy $-{\eps}_n$ (see for example Ref.~\cite{deGennes89}).

In addition, on the tight-binding lattice, if the chemical potential $\mu=0$ in the
above BdG Hamiltonian (\ref{bcshamnambu}), then there is a {\it particle-hole} symmetry
in the following sense: if $(u_n(\br), v_n(\br))$ is a solution with energy ${\eps}_n$,
then there is always another solution $e^{i\pi(r_x+r_y)}(u_n(\br),v_n(\br))$ with energy $-{\eps}_n$. Thus we can choose:
\begin{equation}
\left( \begin{array}{c} u^{(-)}_n(\br) \\
v^{(-)}_n(\br) \end{array} \right)=e^{i\pi(r_x+r_y)}
\left( \begin{array}{c} u^{(+)}_n(\br) \\
v^{(+)}_n(\br) \end{array} \right),
\label{particlehole}
\end{equation}
where $+$ ($-$) corresponds to a solution with positive (negative) energy eigenvalue.
We will refer to this as particle hole transformation $\hat{P}_H$.
\subsection{Franz-Te\v{s}anovi\'c Transformation and Translation Symmetry}
In order to elucidate another important symmetry of the Hamiltonian (\ref{bcshamnambu}),
we follow FT \cite{ft,vmft} and perform a ``bipartite'' singular gauge
transformation on the Bogoliubov-de~Gennes Hamiltonian (\ref{bdg}),
\begin{equation}
\hat{H}_0 \to U^{-1} \hat{H}_0 U, \ \ \
U=\left( \begin{array}{cc}
e^{i\phi_e({\bf r})}      & 0 \\
0  & e^{-i\phi_h({\bf r})}
\end{array} \right),
\label{u2}
\end{equation}
where $\phi_e({\bf r})$ and $\phi_h({\bf r})$ are two auxiliary
vortex phase functions satisfying
\begin{equation}
\phi_e({\bf r})+\phi_h({\bf r})=\phi({\bf r}).
\label{con1}
\end{equation}
This transformation eliminates the phase of the order
parameter from the pairing term of the Hamiltonian. The phase fields
$\phi_e({\bf r})$ and $\phi_h({\bf r})$ can be chosen in a way that avoids
multiple valuedness of the wavefunctions. The
way to accomplish this is to assign the singular part of the phase field
generated by any given vortex to either $\phi_e({\bf r})$ or
$\phi_h({\bf r})$, but not both.
Physically, a vortex assigned to $\phi_e({\bf r})$ will be seen
by electrons and be invisible to holes, while vortex assigned to
$\phi_h({\bf r})$ will be seen by holes and be invisible to electrons.
\begin{figure}[t]
\epsfxsize=6.5cm
\hfil\epsfbox{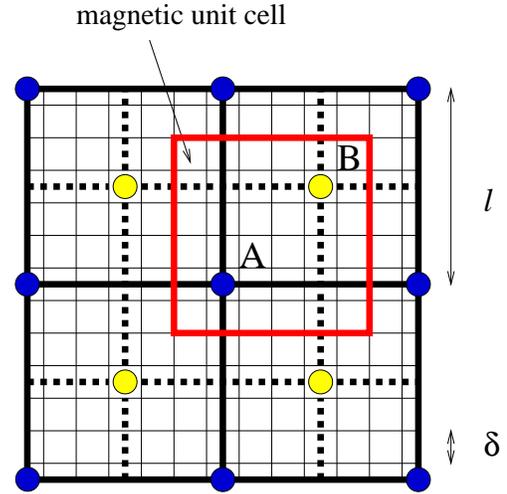}\hfill
\caption{Example of $A$ and $B$ sublattices for the square vortex arrangement.
 The underlying tight-binding lattice, on which the electrons and holes
are allowed to move, is also indicated.}
\label{transf}
\end{figure}
For periodic Abrikosov vortex array, we implement the above transformation
by dividing vortices into
two groups $A$ and $B$, positioned at $\{\br_i^A\}$ and $\{\br_i^B\}$
respectively (see Fig.~\ref{transf}). We then define two phase fields
$\phi^A({\bf r})$ and  $\phi^B({\bf r})$ such that
\begin{equation}
\nabla\times\nabla\phi^\alpha({\bf r}) = 2\pi\hat{z}\sum_i
\delta ({\bf r} -{\bf r}_i^\alpha), \ \ \ \alpha=A,B,
\label{del}
\end{equation}
and identify $\phi_e=\phi^A$ and $\phi_h=\phi^B$.
On the tight-binding lattice the transformed Hamiltonian becomes
\begin{multline}
\hat{H}_N=\sum_{\bdelta} \left\{ \sigma_3 \bigl{(}
-t e^{i\int_{\br}^{\br+\bdelta}({\bf a}-\sigma_3 {\bf v})\cdot d\bl}e^{i\bdelta \cdot \bp}-\mu\bigr{)}\right.\\
\left.+\sigma_1 \Delta_{0}\eta_{\bdelta}e^{i\int_{{\bf r}}^{\br+\bdelta}{\bf a}\cdot d\bl}e^{i\bdelta \cdot \bp}
\right\}
\label{calv}
\end{multline}
where
\begin{equation}
{\bf v}=\frac{1}{2} \nabla \phi - \frac{e}{c}\bA; \ \
{\bf a}=\frac{1}{2}(\nabla \phi^A - \nabla \phi^B),
\end{equation}
$\sigma_1$ and $\sigma_3$ are Pauli matrices operating in Nambu space,
and the sum is again over the nearest neighbors.
Note that the integrand of Eq.\ (\ref{calv}) is proportional to the
superfluid velocities
\begin{equation}
{\bf v}_s^\alpha=\frac{1}{m^*}(\nabla\phi^\alpha-\frac{e}{c}{\bf A}), \ \ \alpha=A,B.
\label{vsab}
\end{equation}
and is therefore explicitly gauge invariant as are the off-diagonal pairing terms.

From the perspective of quasiparticles ${\bf v}_s^A$ and ${\bf v}_s^B$
can be thought of as
{\em effective} vector potentials acting on electrons and holes respectively.
Corresponding effective magnetic field seen by the quasiparticles is
${\bf B}_{\rm eff}^\alpha=- \frac{m^*c}{e}(\nabla\times{\bf v}_s^\alpha)$, and can be
expressed using Eqs.\ (\ref{del}) and (\ref{calv}) as
\begin{equation}
{\bf B}_{\rm eff}^\alpha=\bB-\phi_0\hat{z}\sum_i
\delta ({\bf r} -{\bf r}_i^\alpha), \ \ \ \alpha=A,B,
\label{beff}
\end{equation}
where $\bB=\nabla\times\bA$ is the physical magnetic field and
$\phi_0=hc/e$ is the flux quantum. We observe that quasi-electrons and
quasi-holes propagate in the effective field which consists of
(almost) uniform
physical magnetic field $\bB$ and an array of opposing delta function
``spikes'' of unit fluxes associated with vortex singularities. The latter
are different for electrons and holes.
As discussed in \cite{ft,vmft} this choice guarantees that the effective magnetic field
vanishes on average, i.e. $\langle{\bf B}_{\rm eff}^\alpha\rangle=0$
since we have precisely one flux spike
(of $A$ and $B$ type) per flux quantum of the physical magnetic field.
Flux quantization guarantees that the right hand side of Eq.\
(\ref{beff}) vanishes when averaged over a vortex lattice unit cell containing
two physical vortices. It also implies that there must be equal numbers of
$A$ and $B$ vortices in the system.

The essential advantage of the choice with vanishing
$\langle{\bf B}_{\rm eff}^\alpha\rangle$ is
that  ${\bf v}_s^A$ and ${\bf v}_s^B$ can be chosen periodic in space
with periodicity of the magnetic unit cell containing an
integer number of electronic
flux quanta $hc/e$. Notice that vector potential of a field that does not
vanish on average can never be periodic in space. Condition
$\langle{\bf B}_{\rm eff}^\alpha\rangle=0$
is therefore crucial in this respect.
The singular gauge transformation (\ref{u2}) thus
maps the original Hamiltonian
of fermionic quasiparticles in finite magnetic field
onto a new Hamiltonian
which is formally in zero average field and has
only "neutralized" singular phase windings in the off-diagonal components.

The resulting new Hamiltonian now commutes with translations spanned by
the magnetic unit cell i.e.
\begin{equation}
[\hat{T}_{{\bf R}},\hat{H}_N]=0,
\end{equation}
where  the translation operator $\hat{T}_{{\bf R}}= \exp(i\bR \cdot \bp)$.
We can therefore label eigenstates with a ``vortex'' crystal momentum
quantum number $\bk$ and use the familiar Bloch states as
the natural basis for the eigen-problem.
In particular we seek the eigensolution of the BdG
equation $\hat{H}_N\psi=\epsilon\psi$ in the Bloch form
\begin{equation}
\psi_{n\bk}(\br) = e^{i\bk\cdot\br}\Phi_{n\bk}(\br)=
e^{i\bk\cdot\br}\left( \begin{array}{c} U_{n\bk}(\br) \\
V_{n\bk}(\br) \end{array} \right),
\label{bloch}
\end{equation}
where $(U_{n\bk},V_{n\bk})$ are periodic on the corresponding unit cell,
$n$ is a band
index and $\bk$ is a crystal wave vector. Bloch
wavefunction  $\Phi_{n\bk}(\br)$ satisfies the ``off-diagonal'' Bloch
equation $\hat{H}(\bk)\Phi_{n\bk}=\epsilon_{n\bk}\Phi_{n\bk}$ with the
Hamiltonian of the form
\begin{equation}
\hat{H}(\bk)=e^{-i\bk\cdot\br}\hat{H}_N e^{i\bk\cdot\br}.
\label{kham}
\end{equation}
Note, that the dependence on $\bk$, which is bounded to
lie in the first Brillouin zone, is continuous.
This will become important when topological properties of
spin transport are discussed in the Section III C.

\section{Spin Conductivity}
Within the framework of linear-response theory \cite{mahan},
spin dc conductivity can be related to the spin current-current retarded correlation
function $D^{R}_{\mu \nu}$ through :
\begin{equation}
\sigma^s_{\mu \nu}=\lim_{\Omega \to 0}
\lim_{q_1,q_2 \to 0} -\frac{1}{i\Omega}
\biggr{(}D^{R}_{\mu \nu}(q_1, q_2,\Omega)-D^{R}_{\mu \nu}(q_1, q_2,0) \biggl{)}.
\label{sigma}
\end{equation}
The retarded correlation function $D^R_{\mu\nu}(\Omega)$ can in turn be related to the
the Matsubara finite temperature correlation function
\begin{equation}\label{greenspin}
D_{\mu\nu}(i\Omega)=-\int_0^{\beta}
e^{i\tau\Omega}
\bra T_{\tau}j_{\mu}^s(\tau) j_{\nu}^s(0)\ket d\tau
\end{equation}
as
\begin{equation}
\lim_{q_1,q_2 \to 0} D^R_{\mu\nu}(q_1,q_2,\Omega)=D_{\mu\nu}(i\Omega \rightarrow \Omega + i0).
\end{equation}
In the Eq. (\ref{greenspin}) the spatial average of the spin current
${\bf j}^s(\tau)$ is implicit,
since we are looking for dc response of spatially inhomogeneous system.
In the next section we derive the spin current and evaluate the above formulae.

\subsection{Spin Current}
In order to find the dc spin conductivity, we must first find the spin current.
More precisely, since we are looking only for the spatial average of
the spin current ${\bf j}^s(\tau)$ we just need its $\bk \rightarrow 0$ component.
In direct analogy with the $\bB=0$ situation \cite{durst},
we can define the spin current by the continuity equation:
\begin{equation}
\dot{\rho^s}+\nabla \cdot {{\bf j}^s}=0
\label{continuityspin}
\end{equation}
where
$\rho^s=\frac{\hbar}{2} (\psidag_{\uparrow x}\psi_{\uparrow x}-
\psidag_{\downarrow x}\psi_{\downarrow x} ) $ is the spin density
projected onto z-axis.
We can then use equations of motion for the $\psi$ fields (\ref{eqmot})
and compute the current density ${\bf j}^s$ from (\ref{continuityspin}).

In the limit of $q \rightarrow 0$ the spin current can be written as (see Appendix B)
\begin{equation}
j^s_{\mu}=\frac{\hbar}{2}\Psi^{\dagger}V_{\mu}\Psi,
\label{spincurrent}
\end{equation}
where the Nambu field $\Psi^{\dagger}=(\psi^{\dagger}_{\uparrow}, \psi_{\downarrow})$
and the generalized velocity matrix operator $V_{\mu}$ satisfies the following commutator
identity
\begin{equation}
V_{\mu} =\frac{1}{i \hbar} [x_{\mu},\hat{H}_0 ].
\label{velocity}
\end{equation}
The equation (\ref{velocity}) is a direct restatement of the fact
that spin can be transported by diffusion,
i.e. it is a good quantum number in a superconductor,
and that the average velocity of its propagation is
just the group velocity of the quantum mechanical wave.

In the clean limit, the transverse spin conductivity
$\sigma^s_{xy}$ defined in Eq. (\ref{sigma}) is
\begin{equation}
\sigma^s_{xy}(T)=\frac{\hbar^2}{4i}\sum_{m,n}
(f_{n}-f_{m})\frac{V_y^{m n} V_x^{n m}}
{(\eps_{n}-\eps_{m}+i0)^2},
\label{spinsigmaxyT}
\end{equation}
where $f_{m}=\bigl{(}1+\exp(\beta \eps_n)\bigr{)}^{-1}$
is the Fermi-Dirac distribution function evaluated at energy $\eps_{m}$.
For details of the derivation see Appendix B.
The indices $m$ and $n$ label quantum numbers of particular states. The matrix
elements $V_{\mu}^{m n}$ are
\begin{equation}\label{velmatel}
V_{\mu}^{m n}=
\bigl{\langle}m \big{|}V_{\mu}\big{|}n \bigr{\rangle}=
\int d\bx \;
({u}^{\ast}_{m},{v}^{\ast}_{m})
V_{\mu}
\left( \begin{array}{c}
u_{n}\\
v_{n}
\end{array} \right)
\end{equation}
where the particle-hole wavefunctions $u_{m},v_{n}$ satisfy
the Bogoliubov-deGennes equation (\ref{bdg}).
Note that unlike the longitudinal dc conductivity,
transverse conductivity is well defined even in the absence of
impurity scattering. This demonstrates the fact that the transverse conductivity
is {\em not} dissipative in origin. Rather, as will be discussed in
Section III C, its nature is topological.

In the limit of $T \rightarrow 0$ the expression (\ref{spinsigmaxyT})
for $\sigma^s_{xy}$ becomes
\begin{equation}
\sigma^s_{xy}=\frac{\hbar^2}{4i}\sum_{\eps_{m}<0<\eps_{n}}
\frac{V_x^{m n} V_y^{n m}-V_y^{m n} V_x^{n m} }
{(\eps_{m}-\eps_{n})^2}.
\label{spinsigmaxyT0}
\end{equation}
The summation extends over all states
below and above the Fermi energy which, by the nature of the superconductor,
is automatically set to zero.
\subsection{Vanishing of the Spin Conductivity at Half Filling ($\mu=0$)}
It is useful to contrast the {\it semiclassical} approach
with the full quantum mechanical treatment of transverse spin conductivity.
In semiclassical analysis the starting unperturbed Hamiltonian
is usually defined in the {\it absence} of magnetic field $\bB$.
One then assumes semiclassical
dynamics and no inter-band transitions. In this picture, if there is
particle-hole symmetry in the original ($\bB=0$) Hamiltonian, then
there will be no transverse spin (thermal) transport, since the number of carriers
with a given spin (energy) will be the same in opposite directions. In this context,
similar argument was put forth in Ref.~\cite{simonlee}.
However, the problem of a d-wave superconductor is not so straightforward.
As pointed out in Ref.\cite{vmft},
in the nodal (Dirac fermion) approximation, the vector
potential is solely due to the superflow while the uniform magnetic field
enters as a Doppler shift i.e. Dirac scalar potential. Semiclassical analysis
must then be started from this vantage point and the above conclusions
are not straightforward, since the quasiparticle motion
is irreducibly quantum mechanical.

Here we present an argument for the full quantum mechanical problem,
without relying on the semiclassical analysis.
We show that spin conductivity tensor (\ref{spinsigmaxyT})
vanishes at $\mu=0$ due to particle hole symmetry (\ref{particlehole}).
First note that the Fermi-Dirac distribution function satisfies
$f(\eps)=1-f(-\eps)$. Therefore, the factor $f_m-f_n$ changes sign
under the particle-hole transformation $\hat{P}_H$ (\ref{particlehole}) while
the denominator $(\eps_m-\eps_n)^2$ clearly remains unchanged.
In addition, each of the matrix elements $V^{mn}_{\mu}$ changes sign
under $\hat{P}_H$. Thus the double summation over all states in Eq. (\ref{spinsigmaxyT})
yields zero.

Consequently the spin transport vanishes for a clean strongly type-II
BCS d-wave superconductor on a tight binding lattice at half filling.
Due to Wiedemann-Franz law, which we derive in the next section,
thermal Hall conductivity also vanishes at half filling at
sufficiently low temperatures.
Note that this result is independent of the vortex
arrangement i.e. it holds even for disordered vortex array and
does not rely on any approximation regarding
inter- or intra- nodal scattering.

\subsection{Topological Nature of Spin Hall Conductivity at T=0}
In order to elucidate the topological nature of $\sigma^s_{xy}$, we
make use of the translational symmetry discussed in Section II B and
formally assume that the vortex arrangement is periodic. However,
the detailed nature of the vortex lattice will not be specified
and thus any vortex arrangement is allowed within the magnetic unit cell.
The conclusions we reach are therefore quite general.

We will first rewrite the velocity matrix elements $V^{mn}_{\mu}$ using the singularly
gauge transformed basis as discussed in Section II B.
Inserting unity in the form of the FT gauge transformation (\ref{u2})
\begin{equation}
V^{mn}_{\mu}=\bigl{\langle}m \big{|}V_{\mu}\big{|}n \bigr{\rangle}=
\bigl{\langle}m \big{|}U\; U^{-1} V_{\mu}U\; U^{-1}\big{|}n \bigr{\rangle}.
\end{equation}
The transformed basis states $U^{-1}\big{|}n \bigr{\rangle}$ can now be written
in the Bloch form as  $e^{i\bk\cdot\br}\big{|}n_{\bk} \bigr{\rangle} $ and therefore the
matrix element becomes
\begin{equation}
V^{mn}_{\mu}=
\bigl{\langle}m_{\bk}\big{|}e^{-i\bk\cdot\br} U^{-1}
V_{\mu}U e^{i\bk\cdot\br} \big{|}n_{\bk} \bigr{\rangle}=
\bigl{\langle}m_{\bk}\big{|}V_{\mu}(\bk)\big{|}n_{\bk} \bigr{\rangle}.
\end{equation}
We used the same symbol $\bk$ for both bra and ket because the crystal momentum
in the first Brillouin zone is conserved.
The resulting velocity operator can now be simply expressed as
\begin{equation}
V_{\mu}(\bk) =\frac{1}{\hbar} \frac{\partial \hat{H}(\bk)}{\partial k_{\mu}},
\label{kvel}
\end{equation}
where $\hat{H}_0(\bk)$ was defined in (\ref{kham}).
Furthermore the matrix elements of the partial derivatives of $\hat{H}(\bk)$ can be
simplified according to
\begin{multline}
\bigl{\langle}m_{\bk} \big{|}\frac{\partial \hat{H}({\bk})}{\partial k_{\mu}}
\big{|}n_{\bk} \bigr{\rangle}=
(\eps^{n}_{\bk}-\eps^{m}_{\bk})
\bigl{\langle}m_{\bk} \big{|}\frac{\partial n_{\bk}}{\partial k_{\mu}} \bigr{\rangle}\\
=-(\eps^{n}_{\bk}-\eps^{m}_{\bk})
\bigl{\langle}\frac{\partial m_{\bk}}{\partial k_{\mu}} \big{|} n_{\bk} \bigr{\rangle}
\label{matelements},
\end{multline}
for $m \neq n$.
Utilizing Eqs. (\ref{kvel}) and (\ref{matelements}), Eq. (\ref{spinsigmaxyT0})
for $\sigma^s_{xy}$ can now be written as
\begin{multline}
\sigma^{s}_{xy}=\frac{\hbar}{4i}\int \frac{d \bk}{(2\pi)^2} \sum_{\eps^{m}<0<\eps^{n}}
\bigl{\langle}\frac{\partial m_{\bk}}{\partial k_x} \big{|} n_{\bk} \bigr{\rangle}
\bigl{\langle}n_{\bk} \big{|}\frac{\partial m_{\bk}}{\partial k_y} \bigr{\rangle}-\\
\bigl{\langle}\frac{\partial m_{\bk}}{\partial k_y} \big{|} n_{\bk} \bigr{\rangle}
\bigl{\langle}n_{\bk} \big{|}\frac{\partial m_{\bk}}{\partial k_x} \bigr{\rangle}.
\end{multline}
The identity
$\sum_{\eps^{m}_{\bk}<0<\eps^{n}_{\bk}}(|m_{\bk} \rangle \langle m_{\bk}| + |n_{\bk} \rangle \langle n_{\bk}|)=1$,
can be further used to simplify the above expression to read
\begin{equation}
\sigma_{xy}^{s,m}=
\frac{\hbar}{8\pi}\frac{1}{2\pi i}\int d\bk
\biggl{(}
\bigbra \frac{\partial m_{\bk}}{\partial k_x} \bigg{|}
\frac{\partial m_{\bk}}{\partial k_y} \bigket-
\bigbra \frac{\partial m_{\bk}}{\partial k_y} \bigg{|}
\frac{\partial m_{\bk}}{\partial k_x} \bigket
\biggr{)}
\label{sigmaxyalpha}
\end{equation}
where $\sigma_{xy}^{s,m}$ is a contribution to the spin Hall conductance
from a completely filled band $m$, well separated from the rest of the spectrum.
Therefore the integral extends over the
entire magnetic Brillouin zone that is topologically a two-torus $T^2$.
Let us define a vector field $\hat{A}$ in the magnetic Brillouin zone as
\begin{equation}
\hat{A}(\bk)=
\langle m_{\bk}|{\bf \nabla}_{\bk}| m_{\bk} \rangle,
\label{A}
\end{equation}
where ${\bf \nabla}_{\bk}$ is a gradient operator in the $\bk$ space.
From (\ref{sigmaxyalpha}) this contribution becomes
\begin{equation}
\sigma_{xy}^{s,m}
=\frac{\hbar}{8\pi}\frac{1}{2\pi i}
\int d\bk[{\bf \nabla}_{\bk} \times \hat{A}(\bk)]_{z},
\label{sigmatop}
\end{equation}
where $[]_{z}$ represents the third component of the vector.
The topological aspects of the quantity in (\ref{sigmatop})
were extensively studied in the context of integer quantum Hall effect
(see e.g. \cite{kohmoto}) and it is a well known fact that
\begin{equation}
\frac{1}{2\pi i} \int d\bk[{\bf \nabla}_{\bk} \times \hat{A}(\bk)]_{z}=C_1
\end{equation}
where $C_1$ is a first Chern number that is an integer.
Therefore, a contribution of each filled band to $\sigma_{xy}^s$ is
\begin{equation}
\sigma_{xy}^{s,m}=\frac{\hbar}{8\pi}N
\end{equation}
where N is an integer.
The assumption that the band must be separated
from the rest of the spectrum can be relaxed.
If two or more fully filled
bands cross each other the sum total of their
contributions to spin Hall conductance is
quantized even though nothing guarantees the
quantization of the individual contributions.
The quantization of the total spin Hall conductance
requires a gap in the single particle
spectrum at the Fermi energy.
As discussed in the Section V, the general single particle spectrum
of the d-wave superconductor in the vortex
state with inversion-symmetric vortex lattice
is gapped and therefore the
quantization of $\sigma^s_{xy}$ is guaranteed.

\section{Thermal Conductivity}
Before discussing the nature of the quasi-particle spectrum, we will establish
a Wiedemann-Franz law between spin conductivity and thermal conductivity for a d-wave
superconductor.
This relation is naturally expected to hold for a very general system in which
the quasi-particles form a degenerate assembly i.e. it holds even in the presence
of elastically scattering impurities.

Following Luttinger \cite{luttinger}, and Smr\v{c}ka and St\v{r}eda \cite{smst}
we introduce a pseudo-gravitational potential \mbox{$\chi=\bx \cdot {\bf g} /c^2$}
into the Hamiltonian (\ref{bcsnambu}) where ${\bf g}$ is a constant vector.
The purpose is to include a
coupling to the energy density on the Hamiltonian level.
This formal trick allows us to equate statistical
($T\nabla{(1/T)}$) and mechanical (${\bf g}$) forces so that
the thermal current ${\bf j}^Q$, in the long wavelength limit given by
\begin{equation}
{\bf j}^{Q}=L^Q(T) \left(T\nabla\frac{1}{T}-\nabla\chi\right),
\end{equation}
will vanish in equilibrium. Therefore it is enough to consider only the
dynamical force ${\bf g}$ to calculate the phenomenological coefficient
$L^Q_{\mu \nu}$. Note that thermal conductivity $\kappa_{xy}$ is
\begin{equation}
\kappa_{\mu \nu}(T)=\frac{1}{T} L^{Q}_{\mu \nu}(T).
\label{kapLQ}
\end{equation}

When the BCS Hamiltonian $H$ introduced in Eq.(\ref{bcshamiltonian})
becomes perturbed by the pseudo-gravitational field, the resulting Hamiltonian $H_T$
has the form
\begin{equation}
H_T=H + F
\label{hamthermal}
\end{equation}
where $F$ incorporates the interaction with the perturbing field:
\begin{equation}
F=\frac{1}{2}\int d {\bf x} \;\; \Psi^{\dagger}(\bx)\; (\hat{H}_0 \chi +\chi \hat{H}_0) \;\Psi(\bx).
\end{equation}
Since $\chi$ is a small perturbation, to the first order in $\chi$
the Hamiltonian $H_T$ can be written as
\begin{equation}
H_T=\int d {\bf x} \;\; (1+\frac{\chi}{2})\Psi^{\dagger}(\bx)\; \hat{H}_0 \;(1+\frac{\chi}{2})\Psi(\bx)
\end{equation}
i.e. the application of the pseudo-gravitational field results in
rescaling of the fermion operators:
\begin{equation}
\label{rescalingops}
\Psi \rightarrow \tilde{\Psi}=(1+\frac{\chi}{2})\Psi.
\end{equation}

If we measure the energy relative to the Fermi level, the transport of heat
is equivalent to the transport of energy. In analogy with the Section III A,
we define the heat current ${\bf j}^Q$ through diffusion of the energy-density $h_T$.
From conservation of the energy-density the continuity equation follows
\begin{equation}
\label{contineq}
\dot{h}_T+\nabla \cdot {{\bf j}^Q}=0.
\end{equation}
In the limit of $q \rightarrow 0$ the thermal current is
\begin{equation}
j_{\mu}^Q=
\frac{i}{2}
\left(\tilde{\Psi}^{\dagger} V_{\mu}\dot{\tilde{\Psi}}-
\dot{\tilde{\Psi}}^{\dagger} V_{\mu}\tilde{\Psi}\right).
\end{equation}
For details see Appendix C.
Note that the quantum statistical average of the current has two contributions,
both linear in $\chi$,
\begin{equation}
\bra j^Q_{\mu} \ket = \bra j^Q_{0 \mu} \ket+\bra j^Q_{1 \mu} \ket {\equiv}
-(K^Q_{\mu \nu} + M^Q_{\mu \nu}) \partial_{\nu}\chi.
\end{equation}
The first term is the usual Kubo contribution to $L^Q_{\mu\nu}$ while the second term is
related to magnetization of the sample \cite{girvin} for transverse components of $\kappa_{\mu \nu}$
and vanishes for the longitudinal components.
In Appendix C we show that at $T=0$ the term related to magnetization
cancels the Kubo term and therefore
the transverse component of $\kappa_{\mu \nu}$ is zero at $T=0$.
To obtain finite temperature response, we perform Sommerfeld expansion and derive
Wiedemann-Franz law for spin and thermal Hall conductivity.

As shown in the Appendix C
\begin{equation}
L^{Q}_{\mu\nu}(T)=- \biggl{(}\frac{2}{\hbar}\biggr{)}^2
\int \! d\xi \;\xi^2 \frac{d f(\xi)}{d \xi} \tilde{\sigma}^{s}_{\mu\nu}(\xi)
\label{LQT}
\end{equation}
where
\begin{equation}
\tilde{\sigma}^s_{xy}(\xi)=\frac{\hbar^2}{4i}\sum_{\eps_{m}<\xi<\eps_{n}}
\frac{V_x^{m n} V_y^{n m}-V_y^{m n} V_x^{n m} }
{(\eps_{m}-\eps_{n})^2}.
\label{spinsigmaxyksi}
\end{equation}
Note that $ \tilde{\sigma}^{s}_{\mu \nu}(\xi=0)=\sigma^s_{\mu\nu}(T=0)$.
For a superconductor at low temperature the derivative of the Fermi-Dirac distribution
function is
\begin{equation}
-\frac{d f(\xi)}{d \xi}=\delta(\xi)+\frac{\pi^2}{6}(k_B T)^2
\frac{d^2}{d\xi^2}\delta(\xi)+ \cdots
\label{FermiDirac}
\end{equation}
Substituting (\ref{FermiDirac}) into (\ref{LQT}) we obtain
\begin{equation}
L^{Q}_{\mu \nu}(T)=\frac{4\pi^2}{3\hbar^2}(k_B T)^2
\sigma^s_{\mu\nu},
\end{equation}
where $\sigma^{s}_{\mu\nu}$ is evaluated at $T=0$.
Finally, using (\ref{kapLQ}), in the limit of $T \rightarrow 0$
\begin{equation}
\kappa_{\mu \nu}(T)=\frac{4\pi^2}{3}\biggl{(}\frac{ k_B}{\hbar}\biggr{)}^2 T
\sigma^s_{\mu \nu}.
\label{wiedfranz}
\end{equation}
We recognize the Wiedemann-Franz law for the spin and thermal conductivity in the
above equation. As mentioned, this relation is quite general in that it is
independent of the spatial arrangement of the vortex array or elastic impurities.
Thus, quantization of the transverse spin conductivity $\sigma^s_{xy}$ implies quantization
of $\kappa_{xy}/T$ in the limit of  $T \rightarrow 0$.

\section{Quasiparticle Spectrum and Quantized Conductivity}
General features of the quasi-particle spectrum can be understood on the
basis of symmetry alone. Since the time-reversal symmetry is broken, the
Bogoliubov-de~Gennes Hamiltonian $H_0$ (\ref{bdg}) must be, in general, complex.
According to the ``non-crossing'' theorem of
von Neumann and Wigner \cite{noncrossing}, a complex Hamiltonian can have
degenerate eigenvalues unrelated to symmetry only if there are three parameters
which can be varied simultaneously.

Since the system is two dimensional, with the vortices arranged on the lattice,
there are two parameters in the Hamiltonian $\hat{H}(\bk)$ (\ref{kham}): vortex crystal momenta
$k_x$ and $k_y$ which vary in the first Brillouin zone. Therefore, we should not expect any
degeneracy to occur, {\it in general}, unless there is some symmetry which protects it.
Away from half-filling ($\mu \neq0$) and with unspecified arrangement of vortices in the magnetic
unit cell there is not enough symmetry to cause degeneracy. There is only {\it global} Bogoliubov-de~Gennes symmetry relating
quasi-particle energy $\eps_{\bk}$
at some point $\bk$ in the first Brillouin zone to $-\eps_{-\bk}$.

In order for every quasiparticle band to be either completely below
or completely above the Fermi energy,
it is sufficient for the vortex lattice to have inversion symmetry.
This can be readily seen by the following argument:
Consider a vortex lattice with inversion symmetry. Then, 
by the very nature of the superconducting
vortex carrying $\frac{hc}{2e}$ flux, there must be even number 
of vortices per magnetic unit cell
and we are then free to choose Franz-Tesanovic labels $A$ and $B$ in such a way
that $\bv^A(-\br)=-\bv^B(\br)$. To see this note that the explicit
form of the superfluid velocities can be written as
\cite{vmft}:
\begin{equation}
{\bf v}_s^\alpha(\br)=\frac{2\pi\hbar}{m^*}\int\frac{d^2k}{(2\pi)^2}
\frac{i\bk\times\hat{z}}{k^2}
\sum_ie^{i\bk\cdot(\br-\br_i^\alpha)},
\label{vsabf}
\end{equation}
where $\alpha=A$ or $B$ and $\br_i^{\alpha}$ denotes the position of the vortex with label $\alpha$.
If the vortex lattice has inversion symmetry then for every $\br_i^A$
there is a corresponding $-\br_i^B$
such that $\br_i^A=-\br_i^B$. Therefore, under space inversion ${\cal I}$
\begin{equation}
{\cal I} \bv^A(\br)=\bv^A(-\br)=-\bv^B(\br).
\end{equation}
Recall that the tight-binding lattice Bogoliubov-de~Gennes Hamiltonian written
in the Bloch basis (\ref{kham}) reads:
\begin{multline}
\hat{H}(\bk)=\sum_{\bdelta} \left\{ \sigma_3 \bigl{(}
-t e^{i\int_{\br}^{\br+\bdelta}({\bf a}-\sigma_3 {\bf v})\cdot d\bl}e^{i\bdelta \cdot (\bk+\bp)}
-\mu\bigr{)} \right.\\
\left.+\sigma_1 \Delta_{0}\eta_{\bdelta}e^{i\int_{{\bf r}}^{\br+\bdelta}{\bf a}\cdot d\bl}e^{i\bdelta
\cdot(\bk+\bp)}
\right\}
\label{calv2}
\end{multline}
where
\begin{equation}
{\bf v}(\br)\equiv\frac{1}{2}\bigl{(}\bv^A(\br)+\bv^B(\br)\bigr{)}; \ \
{\bf a}(\br)\equiv\frac{1}{2}\bigl{(}\bv^A(\br)-\bv^B(\br)\bigr{)}.
\end{equation}
As before $\sigma_1$ and $\sigma_3$ are Pauli matrices operating in Nambu space
and the sum is again over the nearest neighbors.
It can be easily seen that upon applying the space inversion ${\cal I}$ to $\hat{H}(\bk)$
followed by complex conjugation ${\cal C}$ and $i \sigma_2$
we have a symmetry that for every $\eps_{\bk}$ there is $-\eps_{\bk}$, that is:
\begin{equation}
-i\sigma_2 {\cal C I} \;\hat{H}(\bk)\; {\cal I C}i\sigma_2= - \hat{H}(\bk)
\end{equation}
which holds for every point in the Brillouin zone. Therefore, in order for the spectrum {\em not} to be gapped,
we would need band crossing at the Fermi level. But by the non-crossing theorem this cannot
happen {\it in general}. Thus, the generic quasi-particle spectrum of inversion symmetric vortex lattice is gapped.
As was established in the previous section, gapped quasi-particle spectrum implies quantization of the transverse
spin conductivity $\sigma^s_{xy}$ as well as $\kappa_{xy}/T$ for $T$ sufficiently low.

\begin{figure}
\epsfxsize=8.0cm
\hfil\epsfbox{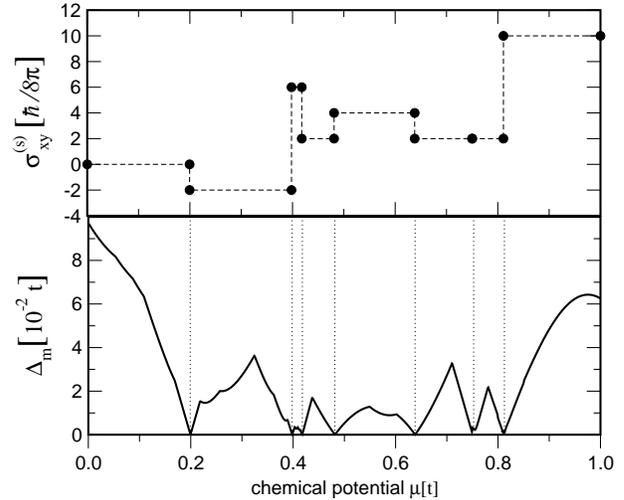}\hfill
\vspace{0.25cm}
\caption{The mechanism for changing the quantized spin Hall conductivity
is through exchanging the topological quanta via (``accidental'') gap closing.
The upper panel displays spin Hall conductivity $\sigma_{xy}^s$
as a function of the chemical potential $\mu$. The lower panel shows
the magnetic field induced gap $\Delta_m$ in the quasiparticle spectrum.
Note that the change in the spin Hall conductivity occurs precisely
at those values of chemical potential at which the gap closes.
Hence the mechanism behind the changes of $\sigma_{xy}^s$ is the
exchange of the topological quanta at the band crossings.
The parameters for the above calculation were:
square vortex lattice, magnetic length $l=4\delta$,
$\Delta=0.1t$ or equivalently the Dirac anisotropy $\alpha_D=10 $.}
\label{sxyvsmu}
\end{figure}

Precisely at half-filling ($\mu=0$) $\sigma^s_{xy}$ must vanish on the basis of particle-hole symmetry
(see Section III B). We can then vary the chemical potential so that $\mu \neq 0$ and break particle-hole
symmetry.
Hence, the chemical potential $\mu$ can serve as the third parameter
necessary for creating the accidental degeneracy,
i.e. at some special values of $\mu^*$ the gap at the Fermi level 
will close (see Fig. \ref{sxyvsmu}).
This results in a possibility of changing the quantized value of $\sigma^s_{xy}$ by
an integer in units of $\hbar/8\pi$
(Fig. \ref{sxyvsmu}).\cite{even} 
By the very nature of the superconducting state,
we achieve {\it plateau} dependence on the chemical potential. 
This is to be contrasted to the
plateaus in the ``ordinary'' integer quantum Hall effect 
which are due to the presence of
disorder. In our case, the system is clean and the plateaus 
are due to the magnetic field
induced gap and superconducting pairing.

\begin{figure}
\epsfxsize=8.0cm
\hfil\epsfbox{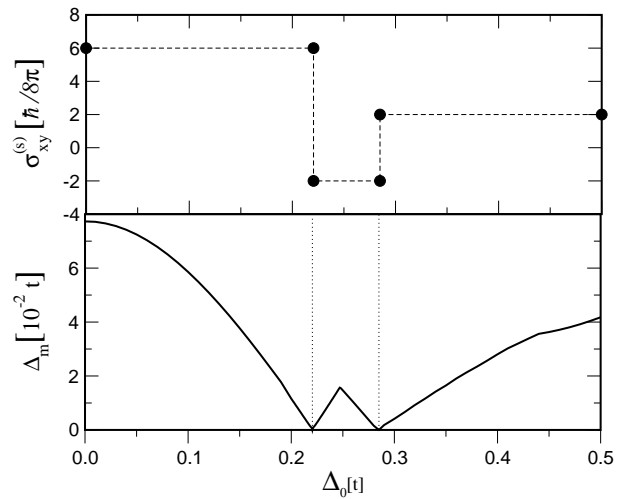}\hfill
\vspace{0.25cm}
\caption{The upper panel displays spin Hall conductivity $\sigma_{xy}^s$
as a function of the maximum superconducting order parameter $\Delta_0$.
The lower panel shows the magnetic field induced gap in the quasiparticle spectrum.
The change in the spin Hall conductivity occurs at those values of $\Delta_0$
at which the gap closes. The parameters for the above calculation were:
square vortex lattice, magnetic length $l=4\delta$, $\mu=2.2t$.}
\label{sxyvsdelta}
\end{figure}

Similarly, we can change the strength of the electron-electron attraction,
which is proportional to the maximum value of the superconducting order parameter $\Delta_0$ while
keeping the chemical potential $\mu$ fixed. Again, as can be seen in Fig. \ref{sxyvsdelta}, at some
special values $\Delta_0^*$ the spectrum is gapless and the quantized Hall conductance
undergoes a transition.

\section{Scaling Functions}
As shown in the previous section, the quasi-particle spectrum of a
d-wave superconductor in vortex state is gapped, in general. Therefore,
at temperatures $T$ which are much less than the  $\Delta_m$ (magnetic field $\bB$ induced gap),
$T$-dependence of thermal conductivity $\kappa_{xy}$ can be determined uniquely.
Moreover, $\bB$-dependence of  $\kappa_{xy}$ comes entirely from the spin conductivity
$\sigma^s_{xy}(\bB)$. That is:
\begin{equation}\label{wiedfranz1}
\kappa_{xy}= \frac{4\pi^2}{3}\biggl{(}\frac{ k_B}{\hbar}\biggr{)}^2 T \sigma^s_{\mu \nu}(B).
\end{equation}
Curiously, if the above equation is naively combined with Simon and Lee scaling \cite{simonlee}
\begin{equation}\label{simonlee}
\kappa_{xy}(T,B)=T^2 F_{xy}\biggl{(}\frac{\sqrt{B}}{T}\biggr{)}.
\end{equation}
the scaling function would be determined up to a proportionality constant $C$:
\begin{equation}
\kappa_{xy}(T,B)=C\; T \sqrt{B}.
\end{equation}
In the recent experiments of Ong {\it et. al.} \cite{ong} this is precisely
the scaling seen in the temperature range up to $\sim 25K$.
Although tempting, the above arguments are unlikely explanation for the experiments since the
measurements are done at rather high temperatures.
\section{Continuum vs. Lattice Theory}
The previous discussion concentrated on the tight-binding formulation of the
problem which is important if the magnetic field is relatively large and if there
is a strong interaction between the underlying ionic lattice and the quasiparticles.
In usual experimental situations, however, the magnetic length is much longer than the inter-ionic
spacing and we would expect that the length-scale associated with the ionic
lattice becomes unimportant at low energies. This leads to a continuum formulation
of the theory (see Section I).
\subsection{Linearized Continuum Theory}
On the basis of the $B=0$ problem,
we expect that the low energy physics is confined in
$k$-space around four nodal points on the
Fermi surface. In order to treat the effect of the
magnetic field and of the vortex lattice
on the low energy properties of the spectrum, we would
ideally like to construct a Hamiltonian which
treats each of the four nodes independently.
A natural way to do this was suggested
by Simon and Lee \cite{simonlee} and later extensively
used by others \cite{ft,marinelli,kallin,vmft}.

As was pointed out in Ref.\cite{vmft}
there is a subtle problem associated with numerical implementation
of the continuum linearized version of the theory.
On the physical grounds any given problem involving charge $e$ (quasi)particle
interacting with a $+\frac{hc}{2e}$ vortex must lead to an identical
eigenvalue spectrum as for a $-\frac{hc}{2e}$ vortex as
long as the wavefunction of the
particle vanishes precisely at the vortex. This is
an exact singular gauge symmetry of our
physical problem and therefore it should be present at
all stages of any approximation.
Extensive numerical study of the linearized
approximation \cite{vmft} pointed out
that this symmetry is weakly violated in the
quasiparticle spectra. The problem
persists irrespective of the techniques chosen to
diagonalize the linearized Hamiltonian i.e. it
is present in real space as well as momentum space representation.
The issue was resolved in Ref.\cite{vmft} by regularizing the problem
on the tight-binding lattice which offered many
advantages, some of which are utilized in this paper.
On the other hand, the concept of
single node physics is lost in a tight
binding formulation as the latter naturally
describes the physics of all four nodes. It would be
desirable to formulate a properly regularized
Hamiltonian describing single node physics only.
From the work of Franz and Tesanovic \cite{ft},
the most natural candidate is a free, anisotropic,
massless Dirac Hamiltonian
\begin{equation}
{\cal H}_0 =\left( \begin{array}{cc}
v_F\hat p_x & v_\Delta\hat p_y \\
v_\Delta\hat p_y & -v_F\hat p_x
\end{array} \right)
\label{FTDirac}
\end{equation}
perturbed by scalar and vector potentials defined in Eq. (\ref{vsab}):
\begin{equation}
{\cal H}' =m\left( \begin{array}{cc}
v_Fv_{sx}^A & \frac{1}{2}v_\Delta(v_{sy}^A-v_{sy}^B) \\
\frac{1}{2}v_\Delta(v_{sy}^A-v_{sy}^B) & v_Fv_{sx}^B
\end{array} \right).
\label{h'}
\end{equation}
$v_F$ and $v_\Delta$ are Fermi and gap velocities respectively \cite{ft}.
We wish to argue that the above Hamiltonian is well defined and does not suffer
from singular gauge asymmetry, provided that suitable boundary conditions and their transformations
are also specified.
The discrepancies encountered in the numerics are believed to result from neglecting
the boundary conditions and the effect of the singular gauge transformation on them.

In the context of high energy physics, a similar problem was studied extensively.
It was found that the Dirac-type equations in the presence of a single Aharonov-Bohm string
require self-adjoint extensions which enlarge the Hilbert space by the wavefunctions
with $r^{-\frac{1}{2}}$ radial behavior. In addition, the wavefunctions
depend on a dimensionless parameter $\Theta$ which specifies the boundary conditions at the core.
A consistent procedure enabling one to construct the needed self-adjoint extension involves the
theory of von Neumann deficiency indices \cite{mreedbsimon,richtmyer}.

The above considerations can be illustrated on a well studied problem of a Dirac
 particle in the field of a single string \cite{sousa,sousajackiw}.
We will restrict ourselves
to the case of a massless Dirac particle and $\frac{hc}{2e}$ flux carried by the string.
The Hamiltonian
\begin{equation}
\label{Dirac}
(i\not\!\partial - e\not\!\!A)\psi = 0
\end{equation}
in this case is axially symmetric and after separation of variables
\begin{equation}
\psi(r,\phi)=
\left(
\begin{array}{c}
\chi^1\\
e^{i\phi}\chi^2
\end{array}
\right)
e^{in\phi}
\end{equation}
one obtains the equation for the radial wavefunctions:
\begin{equation}
\label{DiracRadial}
H_r\chi \equiv \left(
\begin{array}{cc}
0& -i(\frac{d}{dr}+\frac{\nu+1}{r})\\
-i(\frac{d}{dr}-\frac{\nu}{r})&0
\end{array}
\right)
\chi(r)=E\chi(r)
\end{equation}
where $\nu=\frac12+n$ is a half-integer.
The solutions of this equation can be expressed through Bessel functions as
\begin{equation}
\chi_{r}\propto
\left(
\begin{array}{c}
J_{\eps\nu}(|E|r)\\
iJ_{\eps(\nu+1)}(|E|r)
\end{array}
\right)
\end{equation}
where $\eps=\pm1$. The square integrability condition determines sign of $\eps$ for all
$\nu$ {\em except} $\nu=-1/2$. Both choices of $\eps$ result in a wavefunction
diverging as $1/\sqrt{r}$ while still remaining square integrable.
The requirement of regularity for all the solutions at $r=0$ turns out to be too
restrictive and results in numerous pathologies such as
incompleteness of the basis. The problem is solved by using von-Neumann
theory of self-adjoint extensions \cite{sousa,sitenko1}. Since the radial $H_r$ operator in
(\ref{DiracRadial}) defined in the domain of regular functions has
deficiency indices (1,1), one is forced to extend the Hilbert space
by relaxing the condition of regularity and allowing for the wavefunctions \cite{sousa}
\begin{equation}
\label{wfxns}
\chi(r)\propto
\frac{1}{r^{1/2}}
\left(
\begin{array}{c}
i\sin \Theta\\
\strut\\
\cos \Theta
\end{array}
\right).
\end{equation}
The angle $\Theta \in (\pi/4,\pi/4+\pi)$, determined by the details of the vortex core,
parameterizes the self-adjoint extension.

The energy eigenstates for $\nu=-1/2$ are given by
\begin{equation}
\label{sousaeigen}
\left(
\begin{array}{c}
\sin\mu J_{-1/2}(|E|r)+(-1)^n \cos\mu J_{1/2}(|E|r)
\strut\\
\sin\mu J_{1/2}(|E|r)-(-1)^n \cos\mu J_{-1/2}(|E|r)
\end{array}
\right)
\end{equation}
where the parameter $\mu=\Theta$ if $n$ is even, while for $n$ odd $\mu=\pi-\Theta$.
In addition, for some values of $\Theta$ there is a bound state \cite{sousa}.

We encounter a peculiar situation in which the long distance physics governed by
Eq. (\ref{Dirac}) still depends on boundary conditions imposed at the core.
Linearization of the original BdG problem neglects terms ``small'' in conventional sense
which, nevertheless, remain effectively present in $\Theta$ dependence of Eq. (\ref{sousaeigen}) .
Although the linearized equation (\ref{Dirac}) has no length scale,
the core physics, now effectively shrunk to a circle with vanishing radius around the singularity,
manifests itself in the wavefunctions (\ref{wfxns}) with the long power-law tails.
In general, there are only two values of $\Theta$ corresponding to the pure $\delta$-function
magnetic field \cite{tarrach}. However, in the case of other contact interactions
present in the core, the parameter $\Theta$ can be arbitrary \cite{tarrach}.

Various regularizations of the problem were studied  leading to
different choices of the self-adjoint extension \cite{regularizations}.
One possibility is to put an impenetrable cylinder of small but finite radius $\rho$ around a vortex thus
forcing the wavefunctions to vanish on the surface of the cylinder. Such a boundary condition
specification will immediately restore the $\pm \frac{hc}{2e}$ vortex invariance because the
value of the wavefunction at the surface of the cylinder will not transform under singular gauge
transformation. Of course, the true boundary conditions must descend from the original
physical problem, in this case from the self-consistent solution of the full Bogoliubov-de~Gennes
equation including the core region. The quasi-particle wavefunctions do not,
in general, vanish around the vortex \cite{ft98}.
Therefore, not only the quasi-particle wavefunctions
but also the constraints imposed on them must be properly
transformed under singular gauge transformation.
\subsection{Beyond the Linearized Theory}
As pointed out by Simon and Lee \cite{simonlee} and then later by Ye \cite{ye},
linearized theory predicts vanishing transverse conductivities.
In order to include the description of Hall effects,
the curvature terms must be included.
Without regularization, the perturbative analysis of
the curvature terms are far from straightforward.
This has to do with the fast rise of both
wavefunctions and perturbations around the core.
If the regularization is taken into account
and the wavefunctions are forced to vanish at some radius $\rho$
around the core, the new length scale $\rho$ will
break the scale invariance of the Dirac equation.
Moreover, the perturbative analysis of the continuum
equation would predict that the
contribution of each node to $\sigma^s_{xy}$ can be either $0$ or
$\pm \frac{1}{2}$ in units of $\hbar/8\pi$ \cite{ashvin}.
Thus, if all four nodes are included, the only possible values of
$\sigma^s_{xy}$ are $0$ or $\pm 2$.
On the other hand, explicit evaluation of $\sigma^s_{xy}$
via the tight-binding lattice regularization indicates
that a wide range of values for $\sigma^s_{xy}$ is possible
(see Figs. \ref{sxyvsmu} and \ref{sxyvsdelta}). An interesting
question remaining for future study
is how to implement the picture presented in Ref. \cite{ashvin}, of individual
Dirac nodes changing $\sigma^s_{xy}$ by $\pm 1$ through
action of the perturbatively determined mass term as an ``elementary building block'' of
more complicated band crossings.

One way to reach the continuum limit is to make the magnetic length $l$
much longer than the tight binding lattice constant $\delta$.
The drawback associated with the tight-binding regularization of the problem stems from the fact that
the continuum limit is not smooth and unless an exact analytical solution is found
it is complicated to analyze numerically.
The similar situation occurs in a tight-binding Hofstadter problem of a 2D
normal electron gas on the lattice.

\section{conventional superconductors}
Although the previous analysis was focused on unconventional d-wave superconductors, the main results
can be directly generalized to conventional 2D s-wave superconductors
in the vortex state: the transverse spin Hall conductivity $\sigma^s_{xy}$ is quantized in units of $\hbar/8\pi$
and by Wiedemann-Franz law
$\kappa_{xy}=\frac{4\pi^2}{3}\biggl{(}\frac{ k_B}{\hbar}\biggr{)}^2 T\sigma^s_{xy}$ as $T\rightarrow0$.
At low magnetic fields, the quasiparticle spectrum is gapped by the virtue
of the lowest Caroli-Matricon-de~Gennes vortex core bound state having
energy $\sim \Delta^2/\eps_F$. In the vortex lattice the CMdG bound states are extended and form bands.
If the vortices are disordered, the states in the band tails will be localized and the effective quasiparticle
gap will be increased because the localized states do not contribute to transport.
As the magnetic field is increased, the spectrum becomes progressively altered
and at some critical field $\bB^*$ it develops nodal points (see Ref.\cite{dsz}).
The transition between states with two different quantized values of $\sigma^s_{xy}$
happens at the $\bB$ induced gap closings.
Therefore, a series of transitions between different spin Hall states is predicted to occur in the
conventional superconductors as well.

\section{conclusions}
In conclusion, we examined a general problem of 2D type-II superconductors in the
vortex state with inversion symmetric vortex lattice.
The single particle excitation spectrum is typically 
gapped and results in
quantization of transverse spin conductivity
 $\sigma^s_{xy}$ in units of $\hbar/8\pi$ \cite{even}.
The topological nature of this phenomenon is discussed.
The size of the magnetic field induced gap $\Delta_m$ 
in unconventional d-wave superconductors
is not universal and in principle
can be as large as several percent of the maximum superconducting gap $\Delta_0$.
By virtue of the Wiedemann-Franz law, which we 
derive for the d-wave Bogoliubov-de~Gennes equation in the
vortex state, the thermal conductivity
$\kappa_{xy}=\frac{4\pi^2}{3}\biggl{(}\frac{ k_B}{\hbar}\biggr{)}^2 T\sigma^s_{xy}$
as $T \rightarrow 0$. Thus at $T \ll \Delta_m$ the quantization of $\kappa_{xy}/T$
will be observable in clean samples with negligible Lande $g$ factor and with well ordered Abrikosov vortex lattice.
In conventional superconductors, the size of $\Delta_m$ is given by Matricon-Caroli-deGennes
vortex bound states $\sim \Delta^2/\eps_F$.

In real experimental situations, Lande $g$ factor is not necessarily small. In fact it is close to 2
in cuprates \cite{poople}.
The Zeeman effect must therefore be included in the analysis. Nevertheless, detailed numerical examination of the
quasiparticle spectra reveals that the spectrum remains gapped for a wide range of physically realizable
parameters even if $g=2$. Thus, although the Zeeman splitting is a competing effect,
in general it is not strong enough
to prevent the quantization of $\sigma^s_{xy}$ and consequently of $\kappa_{xy}/T$.

We have explicitly evaluated the quantized values of $\sigma^s_{xy}$
on the tight-binding lattice
model of $d_{x^2-y^2}$-wave BCS superconductor in the vortex state
and showed that in principle a wide range of integer values can be
obtained. This should be contrasted with the notion that the effect
of a magnetic field on a d-wave superconductor is solely
to generate a $d+id$ state for the order parameter,
as in that case $\sigma^s_{xy}=\pm2$ in units of $\hbar/8\pi$.
In the presence of a vortex lattice, the situation appears to be more complex.

By varying some external parameter, for instance the strength of
the electron-electron attraction which is proportional to
$\Delta_0$ or the chemical potential $\mu$, the gap
closing is achieved i.e. $\Delta_m=0$.
The transition between two different values of
$\sigma^s_{xy}$ occurs precisely when the gap closes and
topological quanta are exchanged. The remarkable new
feature is the {\em plateau} dependence of $\sigma^s_{xy}$ on the
$\Delta_0$ or $\mu$. It is qualitatively different
from the plateaus in the ordinary integer quantum Hall effect which
are essentially due to disorder. In superconductors,
the plateaus happen in a clean system because the gap in the
quasiparticle spectrum is generated by the
superconducting pairing interactions.

\acknowledgments
We would like to thank A. Vishwanath, A. Durst, M. Franz,
I. Gruzberg, B. Halperin, N. Read and G. Volovik for
useful discussions and comments.
We would especially like to thank A. Vishwanath for sharing
his unpublished work with us. This work was supported
in part by the NSF grant DMR00-94981.

\appendix
\newpage
\section{Green's functions: definitions and identities}

The one-particle Green's function matrix \cite{mahan} is defined as
\begin{equation}\label{greensfxn}
\hatG_{\alpha\beta}(\br_1\tau_1;\br_2\tau_2)
\equiv-\bra T_{\tau} \Psi_{\alpha}(1)\Psidag_{\beta}(2)\ket
\end{equation}
where $T_{\tau}$ denotes imaginary time ordering operator and $\alpha=1,2$
denotes components of a Nambu spinor $ \Psidag(1)$ which is a shorthand for
$\Psidag(\br_1,\tau_1)=\bigl{(}\psi^{\dagger}_{\uparrow}(\br_1,\tau_1), \psi_{\downarrow}(\br_1,\tau_1)\bigr{)}$.

Due to the time independence of the Hamiltonian (\ref{bcshamiltonian}),
the Green's function (\ref{greensfxn}) depends only
on the imaginary time difference $\tau= \tau_1-\tau_2$. Therefore, its Fourier transform is given by
\begin{equation}\label{}
\label{fourier}
\begin{array}{lcl}
\hatG(\br_1,\br_2;i\omega)&=&
\int_0^{\beta}e^{i\omega\tau}\hatG(\br_1,\br_2,\tau)d\tau\\
\strut\\
\hatG(\br_1,\br_2;\tau)&=&
\frac1{\beta}\sum\limits_{i\omega}e^{-i\omega\tau}\hatG(\br_1,\br_2;i\omega).\\
\end{array}
\end{equation}
Here \mbox{$\beta=1/(k_B T)$}, $T$ is temperature, and the fermionic frequency
\mbox{ $\omega=(2l+1)\pi/\beta$, $l \in Z$}.

Using the above relations, it is straightforward to derive the spectral representation
of the following correlation functions between $\Psi$ and its imaginary time derivatives
$\partial_{\tau}\Psi \equiv \Psidot$:
\begin{equation}\label{}
\label{spectr1}
\left\{\!\!\!
\begin{array}{l}
\bra T_{\tau} \Psi_{\alpha}(1)\Psidag_{\beta}(2)\ket\\
\bra T_{\tau} \Psidot_{\alpha}(1)\Psidag_{\beta}(2)\ket\\
\bra T_{\tau} \Psi_{\alpha}(1)\Psidotdag_{\beta}(2)\ket\\
\bra T_{\tau} \Psidot_{\alpha}(1)\Psidotdag_{\beta}(2)\ket\\
\end{array}
\right.
\!\!=-\frac1\beta \sum_{i\omega} e^{-iw\tau}\!\!\int\!\!d\eps \frac{\hatA(\br_1,\br_2;\eps)}{i\omega-\eps}
\left\{\!\!\!
\begin{array}{c}
+1\\
-\eps\\
+\eps\\
-\eps^2
\end{array}
\right.
\end{equation}
The spectral function $\hat{A}(\br_1,\br_2;\eps)$ can be written in terms of
eigenfunctions $\Phi_n(\br)=(u_n(\br),v_n(\br))^T$
of the Bogoliubov-deGennes Hamiltonian $\hH$ in the form:
\begin{equation}\label{spectA}
\hatA_{\alpha\beta}(\br_1,\br_2;\eps)
=\sum\limits_n\delta(\eps-\eps_n)\Phi_{n\alpha}(\br_1)\Phi_{n\beta}^{\dagger}(\br_2),
\end{equation}
where $\eps_n$ is the eigen-energy associated with an eigenstate labeled by the quantum number $n$.
Substituting (\ref{spectA}) into (\ref{spectr1}) we can write the above correlation
functions solely in terms of the eigenfunctions $\Phi_n(\br)$:
\begin{equation}\label{}
\label{correl}
\left\{\!\!\!
\begin{array}{l}
\bra T_{\tau} \Psi_{\alpha}(1)\Psidag_{\beta}(2)\ket\\
\bra T_{\tau} \Psidot_{\alpha}(1)\Psidag_{\beta}(2)\ket\\
\bra T_{\tau} \Psi_{\alpha}(1)\Psidotdag_{\beta}(2)\ket\\
\bra T_{\tau} \Psidot_{\alpha}(1)\Psidotdag_{\beta}(2)\ket\\
\end{array}
\right.
\!\!\!=\!-\frac1\beta \sum_{i\omega,n} e^{-iw\tau}
\frac{\Phi_{n\alpha}(\br_1)\Phi^{\dagger}_{n\beta}(\br_2)}{i\omega-\eps_n}
\left\{\!\!\!
\begin{array}{l}
+1\\
-\eps_n\\
+\eps_n\\
-\eps_n^2
\end{array}
\right.
\end{equation}

In the calculations that follow we will also encounter Matsubara summations over the fermionic
frequencies \mbox{$\omega=(2l+1)\pi/\beta$, $l \in Z$}, of the form:
\begin{equation}\label{matsum}
S_{nm}(i\Omega)=\frac1\beta\sum_{i\omega}\frac1{(i\omega-\eps_n)(i\Omega+i\omega-\eps_m)}
\end{equation}
where \mbox{$\Omega=2\pi k/\beta$, $k \in Z$}, is an outside bosonic frequency.
The sum (\ref{matsum}) can be evaluated using standard techniques (see e.g. \cite{mahan}) and yields:
\begin{equation}\label{freqsum}
S_{nm}(i\Omega)=\frac{f_n-f_m}{\eps_n-\eps_m+i\Omega}.
\end{equation}
Here $f_n$ is a short hand for the Fermi-Dirac distribution function
$f(\eps_n)=\bigl{(}1+\exp(\beta \eps_n)\bigr{)}^{-1}$.

In order to derive spin and thermal currents we will need the explicit form
of the  generalized  velocity operator ${\bf V}$ introduced in
(\ref{velocity}):
\begin{equation}
\label{defnvelocity}
{\bf V}=\left( \begin{array}{cc}
\frac{\bpi}{m} & i \hat{{\bf v}}_{\Delta} \\
i\hat{{\bf v}}^*_{\Delta} & \frac{\bpi^{\ast}}{m}
\end{array} \right)
\end{equation}
where the gap velocity operator is given by
\begin{equation}
i\hat{{\bf v}}_{\Delta}=i{\Delta_{0}} \eta_{\bdelta} \bdelta
e^{i \phi({\bf r})/2}
\bigl{(}\;
e^{i \bdelta \cdot \bp}-e^{-i \bdelta \cdot \bp}
\bigr{)}
e^{i \phi({\bf r})/2}
\label{deltavec}
\end{equation}
and the canonical momentum equals
\begin{equation}
\bpi\!=\!-\frac{i}{2}\bdelta
e^{\frac{i}{\hbar}\int_{{\bf r}}^{{\bf r}
+{\bf \delta}} (\bp-\frac{e}{c}\bA)\cdot d{\bf l}}\;
+ h.c.
\label{pivec}
\end{equation}
The following identities for operators $\hat{{\bf v}}_{\Delta}$ and $\bpi$
will be used in the next section:
\begin{equation}
\label{pinabla}
\!\!\left\{
\begin{array}{ccc}
\bpibar \Psidag \cdot \bpi \Psi&=&i\hbar\nabla \cdot (\Psidag \bpi \Psi)+ \Psidag \bpi^2 \Psi\\
\bpibar \Psidag \cdot \bpi \Psi&=&-i\hbar\nabla \cdot (\bpibar \Psidag \;\; \Psi ) +(\bpibar)^2
\Psidag  \;\; \Psi
\end{array}
\right.
\end{equation}
\begin{equation}\label{deltanabla}
\Psidag \Delta \Psi - \Delta \Psidag \;\;
\Psi=\frac12\nabla\cdot\left(\Psidag \;\; \hat{{\bf v}}_{\Delta}\Psi
-\hat{{\bf v}}_{\Delta}\Psidag \;\; \Psi\right)
\end{equation}
The above equations are straightforward to derive in continuum,
while on the tight-binding lattice
Eqs. (\ref{pinabla}) and (\ref{deltanabla}) imply a
symmetric definition of the lattice
divergence operator.

The identities (\ref{pinabla},\ref{deltanabla}) explicitly ensure that the generalized
velocity operator $V_{\mu}$ is  Hermitian i.e it satisfies the following identity:
\begin{multline}\label{hermvel}
\int \! d\br {\bf V}^*\Psidag(\br)\;\;\Psi(\br) \equiv
\int \! d\br {\bf V}^*_{\alpha\beta}\Psidag_{\beta}(\br)\;\; \Psi_{\alpha}(\br)\\
=\int \! d\br \Psidag(\br)\;\; {\bf V}\Psi(\br).
\end{multline}

\section{Spin current and spin conductivity tensor}
The time derivative of spin density
$\rho_s=\frac{\hbar}{2} (\psidag_{\uparrow}\psi_{\uparrow}-
\psidag_{\downarrow}\psi_{\downarrow} )$
can be written in Nambu formalism as
\begin{equation}
\label{}
\dot{\rho}^s=\frac{i}{\hbar}[H,\rho^s] =\frac{\hbar}2(\Psidotdag\Psi+\Psidag\Psidot)
\end{equation}
Using equations of motion (\ref{eqmot}) together with
the explicit form of $\hH$ operator (\ref{bcshamnambu}) we obtain
\begin{multline}
\dot{\rho}^s=\frac{i}{2}
\left(
\frac{(\pibar)^2}{2m}\Psidag_1\;\; \Psi_1-\Psidag_1\frac{\pi^2}{2m}\Psi_1\right.
+\Deltaop\Psidag_1 \;\; \Psi_2-\Psidag_1 \Deltaop\Psi_2\\
+\Deltaop^*\Psidag_2\;\;\Psi_1-\Psidag_2\;\;\Deltaop^*\Psi_1
\left.-\frac{\pi^2}{2m}\Psidag_2\;\;\Psi_2+\Psidag_2\frac{(\pibar)^2}{2m}\Psi_2
\right).
\end{multline}

Using the identities (\ref{pinabla},\ref{deltanabla}) it is easy to show that
\begin{equation}
\label{}
\dot{\rho}^s=-\frac{\hbar}4\nabla_{\mu}(\Psidag
V_{\mu}\Psi+V^*_{\mu}\Psidag\;\;\Psi)=-\nabla \cdot {\bf j}^s
\end{equation}
where the generalized velocity operator $V_{\mu}$ is defined in Eq. (\ref{defnvelocity}),
and the last equality follows from the continuity equation (\ref{continuityspin})
relating the spin density $\rho^s$ and the spin current ${\bf j}^s$.
Upon spatial averaging and utilizing Eq. (\ref{hermvel}) we find the $q \rightarrow 0$ limit of the spin current
\begin{equation}
\label{}
j^s_{\mu}=\frac{\hbar}2\Psidag V_{\mu}\Psi.
\end{equation}

The evaluation of the spin current-current correlation function
(\ref{greenspin}) is straightforward and yields:
\begin{multline}\label{dspin}
D_{\mu\nu}(i\Omega)=-\frac{\hbar^2}4\int_0^{\beta}
e^{i\tau\Omega}\bigl[V_{\mu}(2)V_{\nu}(4)\times\\
\bra T_{\tau}\Psidag(1)
\Psi(2)\Psidag(3)\Psi(4)\ket
_{\tiny
\begin{array}{l}
1,2\rightarrow(\bx,\tau)\\
3,4\rightarrow(\by,0)
\end{array}
}\bigr]d\tau
\end{multline}
Here $\Psi(1)\equiv \Psi(\br_1,\tau_1)$ and similarly the operator $V_{\mu}(2)$ acts
only on functions of $\br_2$.
Using Wick's theorem, identity (\ref{correl}), and upon spatial averaging
over $\bx$ and $\by$ we obtain
\begin{equation}
\label{}
D_{\mu\nu}(i\Omega)=\frac{\hbar^2}4\sum_{mn} \bra n| V_{\mu}|m \ket \bra m| V_{\nu}
|n\ket S_{nm}(i\Omega).
\end{equation}
The double summation extends over the eigenstates $|m \ket$ and $|n\ket$
of the Hamiltonian (\ref{bcshamnambu}),
and $ S_{mn}(i\Omega)$ is given in Eq. (\ref{freqsum}).
Analytically continuing $i\Omega\rightarrow\Omega+i0$ we finally obtain the expression for
the retarded correlation function:
\begin{equation}
\label{retspin}
D^R_{\mu\nu}(\Omega)=\frac{\hbar^2}4\sum_{mn}  \frac{V_{\mu}^{nm}
V_{\nu}^{mn}}{\eps_n-\eps_m+\Omega+i0} (f_n-f_m).
\end{equation}
where $V^{\mu}_{mn}$ is defined in (\ref{velmatel}) and $f_n$ is a short hand
for the Fermi-Dirac distribution function
$f(\eps_n)=\bigl{(}1+\exp(\beta \eps_n)\bigr{)}^{-1}$.
Finally, we substitute  the last equation into (\ref{sigma}) to obtain
\begin{equation}\label{sigmaappendix}
\sigma^{s}_{\mu\nu}=
\frac{\hbar^2}{4i}\sum_{mn}\frac{V^{nm}_{\mu}V^{mn}_{\nu}}{(\eps_n-\eps_m+i0)^2}(f_n-f_m)
\end{equation}
\section{Thermal current and thermal conductivity tensor}
In order to calculate thermal currents and thermal
conductivity in the magnetic field we introduce
a pseudo-gravitational potential $\chi=\br\cdot{\bf g}/c^2$ \cite{luttinger,obraz,smst,streda}.
This formal procedure is useful because it illustrates that the transverse
thermal response is not given just by Kubo formula, but in addition
it includes corrections related to magnetization. Throughout this section $\hbar=1$.
The pseudo-gravitational potential enters the Hamiltonian, up to linear order in $\chi$, as
\begin{equation}\label{HT}
H_T=\int d {\bf x} \;\; (1+\frac{\chi}{2})\Psi^{\dagger}(\bx)\; \hat{H}_0 \;(1+\frac{\chi}{2})\Psi(\bx),
\end{equation}
where $H_0$ is the Bogoliubov-deGennes Hamiltonian (\ref{bcshamnambu}).
The equations of motion for the fields $\Psi$ thus become
\begin{multline}\label{eqmotT}
i\dot{\Psi}=[\Psi,H_T]=\chifac \hH\chifac \Psi\\=
(1+\chi) \hH \Psi-i\nabla_{\mu}\chi V_{\mu}\Psi.
\end{multline}
The last equality follows from the commutation relation (\ref{velocity}).
(Note that for $\chi \neq 0$ the Eq. (\ref{eqmotT}) differs from Eq. (\ref{eqmot}).
Throughout this section $\Psidot$ will refer to Eq. (\ref{eqmotT}) unless
explicitly stated otherwise).

To find thermal current $\bj^Q$ we start with the continuity equation
\begin{equation}
\label{contineqapp}
\dot{h}_T+\nabla \cdot {{\bf j}^Q}=0.
\end{equation}
The Hamiltonian density $h_T$ follows from Eq. (\ref{HT}) and reads
\begin{multline}
h_T=\frac1{2m^*}\left(
\pibar_{\mu}\tilde{\Psi}^{\dagger}_1\;\; \pi_{\mu}\tilde{\Psi}_1-
\pi_{\mu}\tilde{\Psi}^{\dagger}_2\;\; \pibar_{\mu}\tilde{\Psi}_2
\right)
-\mu\tilde{\Psi}^{\dagger}_{\alpha}\sigma^3_{\alpha\beta}\tilde{\Psi}_{\beta}\\
+\frac12\left(
\tilde{\Psi}^{\dagger}_{\alpha}\Delta_{\alpha\beta}\tilde{\Psi}_{\beta}+
\Delta_{\alpha\beta}\tilde{\Psi}^{\dagger}_{\alpha}\;\;\tilde{\Psi}_{\beta}\right)
\end{multline}
where $\tilde{\Psi}=\chifac \Psi$.
Taking the time derivative of the Hamiltonian density $h_T$ and using
Eqs.(\ref{pinabla}) and equations of motion (\ref{eqmotT}) we obtain
\begin{multline}
\label{ht1}
\dot{h}_T=i[H_T,h_T]=\frac{i}{2m}\nabla_{\mu}\left(
\Psitildadotdag\Pi_{\mu}\;\;\Psitilda
- \Pi^*_{\mu}\Psitildadag\;\;\Psitildadot
\right)\\
+\frac12
\left(
\Psitilda_{\alpha}\Deltaab\Psitildadot_{\beta}
-\Psitildadotdag_{\alpha}\Deltaab\Psitilda_{\beta}
\right)\\
+\frac12
\left(
\Deltaab\Psitildadotdag_{\alpha}\;\;\Psitilda_{\beta}
-\Deltaab\Psitildadag_{\alpha}\;\;\Psitildadot_{\beta}
\right)
\end{multline}
where we introduced matrix operators
\begin{equation}\label{}
\Deltaab=\left(
\begin{array}{cc}
0&\Deltaop\\
\Deltaop^*&0
\end{array}\right),\qquad
\Pi^{\mu}_{\alpha\beta}=\left(
\begin{array}{cc}
\pi_{\mu}&0\\
0&\pi_{\mu}^*
\end{array}\right).
\end{equation}
Here $\Psitildadot=\chifac\Psidot$ and $\pi_{\mu}$ is defined in Eq. (\ref{pivec}).
Finally we use (\ref{deltanabla}) to extract $\bj^{Q}$ from (\ref{ht1}).
Upon spatial averaging and using the Hermiticity of $V_{\mu}$ (\ref{hermvel})
the thermal current $\bj^Q$ reads
\begin{equation}
j_{\mu}^Q=
\frac{i}{2}
\left(\tilde{\Psi}^{\dagger} V_{\mu}\dot{\tilde{\Psi}}-
\dot{\tilde{\Psi}}^{\dagger} V_{\mu}\tilde{\Psi}\right).
\end{equation}
Note that the expression for $\bj^Q$ contains two terms
\begin{equation}\label{}
\bj^{Q}=\bj^{Q}_0+\bj^{Q}_1
\end{equation}
where $\bj^Q_0$ is independent of $\chi$ and $\bj^Q_1$ is linear in $\chi$.
Explicitly:
\begin{equation}\label{}
\bj^{Q}_0(\br)=\frac{1}{2}\Psidag \{{\bf V}, \hat{H}_0\}{\Psi}
\end{equation}
and
\begin{multline}\label{}
j^{Q}_{\mu,1}(\br)=
-\frac{i}4\partial_{\nu}\chi \Psidag\left(V_{\mu}V_{\nu}-V_{\nu}V_{\mu}\right)\Psi\\
+ \frac{\partial_{\nu}\chi}{4}
\Psidag\left((x_{\nu}V_{\mu}+3V_{\mu}x_{\nu})\hH
+\hH(3x_{\nu}V_{\mu}+V_{\mu}x_{\nu})
\right)\Psi
\end{multline}
where $\{a,b \}=ab+ba$.
Analogously to the situation in the normal metal, the thermal average
of $\bj^Q_1$ does not in general vanish in the presence
of the magnetic field \cite{girvin}.

The linear response of the system to the external perturbation can be described by
\begin{multline}
\bra j^Q_{\mu} \ket = \bra j^Q_{0 \mu} \ket+\bra j^Q_{1 \mu} \ket=
-(K_{\mu \nu} + M_{\mu \nu}) \partial_{\nu}\chi \equiv -L^Q_{\mu\nu} \partial_{\nu}\chi
\end{multline}
where
\begin{equation}
K_{\mu \nu}=-\frac{\delta \bra j^Q_{0\mu} \ket}{\delta\; \partial_\nu \chi}=-\lim_{\Omega \to 0}
\frac{P^{R}_{\mu \nu}(\Omega)-P^{R}_{\mu \nu}(0)}{i\Omega}
\label{kappakubo}
\end{equation}
is the standard Kubo formula for a dc response \cite{mahan}, $P^{R}_{\mu \nu}(\Omega)$
being the retarded current-current correlation function, and
\begin{multline}\label{kappamag}
M_{\mu \nu}=-\frac{\delta \bra j^Q_{1\mu} \ket}{\delta\; \partial_\nu
\chi}\\
=-\sum_n  \eps_n f_n
\bra n|\left\{V_{\mu},x_{\nu}\right\}|n\ket
+\sum_n\frac{i}{4} f_n\bra n|[V_{\mu},V_{\nu}]|n\ket
\end{multline}
is a contribution from ``diathermal'' currents \cite{girvin}. Note that the latter
vanishes for the longitudinal response while it remains finite for the transverse
response. As we will show later in this section, at $T=0$ there is an important cancellation
between (\ref{kappakubo}) and (\ref{kappamag}) which renders the thermal conductivity
$\kappa_{\mu\nu}$ well-behaved and prevents the singularity from the temperature
denominator in Eq. (\ref{kapLQ}).

The retarded thermal current-current correlation function $P^{R}_{\mu \nu}(\Omega)$ can
be expressed in terms of the Matsubara finite temperature correlation function
\begin{equation}\label{kubo}
\calP_{\mu\nu}(i\Omega)=-\int_0^{\beta}d\tau
e^{i\Omega\tau}\bigbra T_{\tau}j_{\mu}(\br,\tau)j_{\nu}(\br',0)\bigket
\end{equation}
as
\begin{equation}
P^{R}_{\mu \nu}(\Omega)=\calP_{\mu\nu}(i\Omega \rightarrow \Omega +i0).
\end{equation}
The current $j_{\mu}(\br,\tau)$ in Eq. (\ref{kubo}) is given by
\begin{equation}
\label{jmatsubara}
j_{\mu}(\tau)=\frac12(\Psidag({\tau})V_{\mu}\partial_{\tau}\Psi(\tau)-
\partial_{\tau}\Psi^{\dagger}(\tau)\;\; V_{\mu}\Psi(\tau))
\end{equation}
As pointed out by Ambegaokar and Griffin \cite{ambegaokar} time ordering
operator $T_{\tau}$ and time derivative operators $\partial_{\tau}$ do not
commute and neglecting this subtlety can lead to formally divergent
frequency summations\cite{durst}.
Taking heed of this subtlety, substitution of (\ref{jmatsubara}) into (\ref{kubo})
amounts to
\begin{multline}
\calP_{\mu\nu}(i\Omega)=-\frac14\int_0^{\beta}d\tau
e^{i\Omega\tau}V^{\mu}_{\alpha\beta}(2)V^{\nu}_{\gamma\delta}(4)
\bigbra T_{\tau}\\
\Psidotdag_{\alpha}(1)\Psi_{\beta}(2)\Psidotdag_{\gamma}(3)\Psi_{\delta}(4)
-\Psidotdag_{\alpha}(1)\Psi_{\beta}(2)\Psidag_{\gamma}(3)\Psidot_{\delta}(4)\\
-\Psidag_{\alpha}(1)\Psidot_{\beta}(2)\Psidotdag_{\gamma}(3)\Psi_{\delta}(4)
+\Psidag_{\alpha}(1)\Psidot_{\beta}(2)\Psidag_{\gamma}(3)\Psidot_{\delta}(4)
\bigket,
\end{multline}
where the notation follows Eq. (\ref{dspin}) and Eq. (\ref{eqmot}).
Upon utilizing the identities (\ref{correl}) and (\ref{hermvel}) and performing the
standard Matsubara summation we have
\begin{equation}\label{}
\calP_{\mu\nu}(i\Omega)=\frac14\sum\limits_{nm}
\bra n|V_{\mu}|m\ket  \bra m|V^{\nu}|n\ket
(\eps_n+\eps_m)^2 S_{nm}(i\Omega)
\end{equation}
where $S_{mn}$ is given by Eq. (\ref{freqsum}).
Analytically continuing $i\Omega \rightarrow \Omega+i0$ we  obtain
\begin{equation}\label{retthermal}
\calP_{\mu\nu}^R(\Omega)=\frac14\sum\limits_{nm}
\frac{(\eps_n+\eps_m)^2\; V_{\mu}^{nm}  V_{\nu}^{mn}}{\eps_n-\eps_m+\Omega+i0}
(f_n-f_m).
\end{equation}
Note that the only difference between the  Kubo contribution to
the thermal response (\ref{retthermal}) and the spin response (\ref{retspin}) is
the value of the coupling constant. In the case of thermal response (\ref{retthermal}),
the coupling constant is $(\eps_n+\eps_m)/2$ which is eigenstate dependent,
while in the case of spin response (\ref{retspin}) it is $\hbar/2$ and eigenstate independent.

Using Eq. ({\ref{kappakubo}) we find that the Kubo contribution to the thermal
transport coefficient is given by
\begin{equation}\label{}
K_{\mu\nu}=-\frac{i}{4}\sum_{nm}\frac{(\eps_n+\eps_m)^2}{(\eps_n-\eps_m+i0)^2}V_{\mu}^{nm}V_{\nu}^{mn}(f_n-f_m).
\end{equation}
This can be written as
\begin{equation}\label{}
K_{\mu\nu}=K^{(1)}_{\mu\nu}+K^{(2)}_{\mu\nu}
\end{equation}
where
\begin{equation}\label{k1}
K^{(1)}_{\mu\nu}=-\frac{i}{4}\sum_{nm}\frac{4\eps_n\eps_m}{(\eps_n-\eps_m+i0)^2}
V_{\mu}^{nm}V_{\nu}^{mn}(f_n-f_m)
\end{equation}
and
\begin{equation}\label{k2}
K^{(2)}_{\mu\nu}=-\frac{i}{4}\sum_{nm}V_{\mu}^{nm}V_{\nu}^{mn}(f_n-f_m)
\end{equation}
Similarly, we can separate the ``diathermal'' contribution (\ref{kappamag}) as
\begin{equation}\label{}
M_{\mu\nu}=M^{(1)}_{\mu\nu}+M^{(2)}_{\mu\nu}
\end{equation}
where $M^{(1)}_{\mu\nu}$ and $M^{(2)}_{\mu\nu}$ refer to the first and second
term in (\ref{kappamag}) respectively. Using the completeness
relation $\sum_m |m\ket\bra m|=1$ it is easy to show that
\begin{equation}\label{m2}
M^{(2)}_{\mu\nu}=\frac{i}{4}\sum_{mn} (f_n-f_m) V_{\mu}^{nm}V_{\nu}^{mn}.
\end{equation}

Comparison of (\ref{k2}) and (\ref{m2}) yields $M^{(2)}_{\mu\nu}+K^{(2)}_{\mu\nu}=0$.
Therefore the thermal response coefficient is given by
\begin{equation}
L^Q_{\mu\nu}=K^{(1)}_{\mu\nu}+M^{(1)}_{\mu\nu}.
\end{equation}

Utilizing commutation relationships (\ref{velocity}),
$M^{(1)}_{\mu \nu}$ can be expressed in the  form:
\begin{equation}
\label{m1eta}
M^{(1)}_{\mu\nu}=\int
\eta f(\eta)\Tr\left(\delta(\eta-\hH)(x^{\mu}V^{\nu}-x^{\nu}V^{\mu})\right)d\eta
\end{equation}
where the integral extends over the entire real line.
The rest of the section follows closely Smr\v{c}ka and St\v{r}eda \cite{smst}.
We define the resolvents $G^{\pm}$:
\begin{equation}
G^{\pm}\equiv(\eta\pm i0-\hH)^{-1}
\end{equation}
and operators
\begin{equation}\label{abfunctions}
\begin{array}{l}
A(\eta)=i\Tr\left(V_{\mu}\frac{dG^+}{d\eta}V_{\nu}\delta(\eta-\hH)
-V_{\mu}\delta(\eta-\hH)V_{\nu}\frac{dG^-}{d\eta}\right)\\
B(\eta)=i\Tr\left(V_{\mu}G^+ V_{\nu}\delta(\eta-\hH)
-V_{\mu}\delta(\eta-\hH)V_{\nu}G^-\right)\\
\end{array}
\end{equation}

To facilitate Sommerfeld expansion we note that response coefficients
$K^{(1)}_{\mu\nu}(T)$, $M^{(1)}_{\mu\nu}(T)$,
$\sigma^s_{\mu\nu}(T)$ have generic form
\begin{equation}\label{sommerfeld1}
L(T)=\int f(\eta) l(\eta) d\eta,
\end{equation}
and after integration by  parts
\begin{equation}\label{sommerfeld}
L(T)=-\int \frac{d f}{d\eta} \tL(\eta)d\eta.
\end{equation}
Here $\tL(\xi)$ is defined as
\begin{equation}\label{lxidefn}
\tL(\xi)\equiv\int_{-\infty}^{\xi}  l(\eta)d\eta.
\end{equation}
Note that $L(T\!=\!0)=\tilde{L}(\xi\!=\!0)$ and in particular the spin
conductivity at $T=0$ satisfies
$\sigma^s_{\mu\nu}(T\!=\!0)=\tilde{\sigma}^s_{\mu\nu}(\xi\!=\!0)$.
Identities (\ref{sommerfeld}) and (\ref{lxidefn})
will enable us to express the coefficients at finite temperature through
the coefficients at zero temperature.
For example, it follows from Eq.~(\ref{m1eta}) that
\begin{equation}
\label{m1}
\tilde{M}^{(1)}(\xi)=\int^{\xi}_{-\infty}
\eta\Tr\left(\delta(\eta-\hH)(x^{\mu}V^{\nu}-x^{\nu}V^{\mu})\right)
\end{equation}
and from  Eqs.(\ref{sigmaappendix}, \ref{abfunctions})
\begin{equation}
\label{sig}
\tilde{\sigma}^{s}_{\mu
\nu}(\xi)\!=\frac{1}{4}\int^{\xi}_{-\infty} A({\eta}) d\eta
\end{equation}
Similarly, coefficient $K^{(1)}_{\mu\nu}$ from (\ref{k1}) can be written as
\begin{multline}
K^{(1)}_{\mu\nu}=-i\int d\eta
f(\eta)\eta\sum \delta(\eta-\eps_n) \eps_m\\
\times
\left(
\frac{V^{nm}_{\mu}V^{mn}_{\nu}}{(\eta-\eps_m+i0)^2}-
\frac{V^{mn}_{\mu}V^{nm}_{\nu}}{(\eta-\eps_m-i0)^2}
\right)
\end{multline}
so that

\begin{multline}
\tilde{K}^{(1)}_{\mu\nu}(\xi)\equiv-i\int^{\xi}_{-\infty}d\eta \;\eta
\sum\delta(\eta-\eps_n)\eps_m\\
\times
\left(
\frac{V^{nm}_{\mu}V^{mn}_{\nu}}{(\eta-\eps_m+i0)^2}-
\frac{V^{mn}_{\mu}V^{nm}_{\nu}}{(\eta-\eps_m-i0)^2}
\right)
\end{multline}
Using definitions (\ref{abfunctions}),  $\tilde{K}^{(1)}(\xi)$  can be
expressed as
\begin{equation}
\label{sdef}
\tilde{K}^{(1)}_{\mu\nu}(\xi)=\int_{-\infty}^{\xi} \eta^2 A(\eta)d\eta+
\int^{\xi}_{-\infty}\eta B(\eta)d\eta.
\end{equation}
After integration by parts one obtains
\begin{equation}\label{kxi}
\tilde{K}^{(1)}_{\mu\nu}(\xi)=
\xi^2\!\!\int\limits_{-\infty}^{\xi} \!\!A(\eta)d\eta+
\!\!\int\limits_{-\infty}^{\xi}\!(\eta^2-\xi^2)
\left(A(\eta)-\frac12\frac{dB}{d\eta}\right)d\eta.
\end{equation}
As shown in Ref. (\cite{smst}) the last term in this expression is exactly
compensated by $\tilde{M}^{(1)}(\xi)$:
This becomes evident after noting that definitions in Eq.~(\ref{abfunctions}) imply
\begin{equation}
\label{}
A-\frac12\frac{dB(\eta)}{d\eta}=\frac12
\Tr\left[\frac{d\delta(\eta-\hH)}{d\eta}(x^{\mu}V^{\nu}-x^{\nu}V^{\mu})\right].
\end{equation}
Substituting  the last identity into the Eq.(\ref{kxi}) and integrating
the second term by parts we obtain
\begin{multline}\label{kxi1}
\tilde{K}^{(1)}_{\mu\nu}(\xi)=
\xi^2\!\!\int\limits_{-\infty}^{\xi} \!\!A(\eta)d\eta\\-
\!\!\int\limits_{-\infty}^{\xi}\!\eta
\Tr\left(
\delta(\eta-\hH)(x^{\mu}V^{\nu}-x^{\nu}V^{\mu})
\right)d\eta
\end{multline}
The second term here is equal to $-M^{(1)}_{\mu\nu}$ from
(\ref{m1}). After the cancellation the result simply reads:
\begin{equation}\label{lxivsa}
\tilde{L}^Q_{\mu\nu}(\xi)=
\tilde{K}^{(1)}_{\mu\nu}(\xi)+\tilde{M}^{(1)}_{\mu\nu}(\xi)=
\xi^2\int_{-\infty}^{\xi} A(\eta)d\eta.
\end{equation}
Or, using (\ref{sig})
\begin{equation}\label{}
\tL^{Q}_{\mu\nu}(\xi)= \biggl{(}\frac{2\xi}{\hbar}\biggr{)}^2
\tilde{\sigma}^{s}_{\mu\nu}(\xi).
\label{LQLS}
\end{equation}
Finally, from Eq.(\ref{sommerfeld}) we find
\begin{equation}
\label{}
L^Q_{\mu\nu}(T)= - \frac{4}{\hbar^2}\int \frac{df(\eta)}{d\eta}
\eta^2\tilde{\sigma}^{s}_{\mu\nu}(\eta){d\eta}.
\end{equation}



\begin{references}
\bibitem{harlingen} D. J. Van Harlingen, Rev. Mod. Phys. {\bf 67}, 515 (1995).
\bibitem{millis} J. Orenstein and A. J. Millis, Science {\bf 288}, 468 (2000).
\bibitem{walker} M. B. Walker and M. F. Smith, \prb {\bf 61}, 11285 (2000).
\bibitem{ding} H. Ding {\it et. al.}, \prb {\bf 54}, R9678 (1996).
\bibitem{taillefert1} M. Chiao {\it et. al.}, cond-mat/9910367.
\bibitem{ong} Y. Zhang {\it et. al.} \prl {\bf 86}, 890 (2001).
\bibitem{simonlee} S. H. Simon and P. A. Lee, \prl {\bf 78}, 1548 (1997).
\bibitem{gorkov} L. P. Gor'kov and J. R. Schrieffer, Phys. Rev. Lett. {\bf 80}, 3360 (1998).
\bibitem{anderson} P. W. Anderson, cond-mat/9812063.
\bibitem{gusynin} E. J. Ferrer, V. P. Gusynin, V. de la Incera, hep-ph/0101308.
\bibitem{kopnin} N. B. Kopnin, V. M. Vinokur, \prb {\bf 62}, 9770 (2000).
\bibitem{ft} M. Franz and Z. Te\v{s}anovi\'c, \prl {\bf 84}, 554 (2000).
\bibitem{marinelli} L. Marinelli, B. I. Halperin and S. H. Simon, \prb {\bf 62}
3488 (2000).
\bibitem{Melnikov} A. S. Melnikov, J. Phys. Cond. Matt. {\bf 82}, 4703 (1999).
\bibitem{kallin} D. Knapp, C. Kallin, A. J. Berlinsky, cond-mat/0011053.
\bibitem{vmft} O. Vafek, A. Melikyan, M. Franz, Z. Te\v{s}anovi\'c, \prb {\bf 63},
134509 (2001).
\bibitem{wang1} Y. Wang and A. H. MacDonald, \prb {\bf 52}, R3876 (1995).
\bibitem{kita1} K. Yasui and T. Kita, \prl {\bf 83}, 4168 (1999).
\bibitem{ashvin} We wish to alert the reader to the new
preprint by A. Vishwanath, cond-mat/0104213, where some of the results
reported in our paper have been independently derived. While our approaches
are different, the results appear to be in general agreement
where applicable and the
two papers could be considered complementary. We direct the
reader's attention to an elegant discussion in cond-mat/0104213 of
symmetry properties of quasiparticle spectra which has helped us in our
own work (See Sec. V), and, in particular, to the issue of
protection of Dirac nodes discussed there.
\bibitem{ye} J. Ye, \prl {\bf 86}, 316 (2001).
\bibitem{nielsen} M. Nielsen and P. Hedeg{\aa}rd,
Phys. Rev. B {\bf 51}, 7679 (1995).
\bibitem{senthil} T. Senthil, M. P. A. Fisher,
L. Balents, and C. Nayak, cond-mat/9808001.
\bibitem{volovik} Quantized spin currents in superfluids and superconductors
in different physical situations have also been considered in
G. E. Volovik, JETP Lett. {\bf 66}, 522 (1997) and
G. E. Volovik and V. M. Yakovenko, J. Phys. Cond. Matt. {\bf 1} 5263, (1989).
See also V. M. Yakovenko, \prl {\bf 65}, 251 (1990).
\bibitem{durst} A. C. Durst and P. A. Lee, \prb {\bf 62}, 1270 (2000).
\bibitem{haldane} This quantization is an example of a general class
of problems exhibiting the ``parity anomaly'' as originally discussed
in F. D. M. Haldane, \prl {\bf 61}, 2015 (1988). For application to
spin currents in chiral magnets see
F. D. M. Haldane and D. P. Arovas, \prb {\bf 52}, 4223 (1995). For
various manifestations of such quantized transport in superfluids
and superconductors see Refs. \cite{senthil,volovik}.
\bibitem{read1} For translationally invariant superconductors
with broken time-reversal and parity such topological invariant has
been derived by N. Read and Dmitry Green, \prb {\bf 61} 10267, (2000).
\bibitem{even} In a typical situation, the combined symmetry of
BdG equations and the inversion symmetry of a vortex lattice
conspire to restrict $\sigma_{xy}^s$ to {\em even}
multiples of $\hbar /8\pi$.
\bibitem{morita} For discussion of some general properties of the
BCS-Hofstadter problem in d-wave superconductors see
Y. Morita and Y. Hatsugai, cond-mat/0007067.
\bibitem{read2} In real superconductors, this sequence of transitions
will also be influenced by strong disorder. For spin
QHE in dirty superconductors with broken orbital
time-reversal symmetry see
I. A. Gruzberg, A. W. Ludwig, and N. Read, \prl {\bf 82} 4524 (1999);
V. Kagalovsky, B. Horovitz, Y. Avishai, and J. T. Chalker, \prl {\bf 82}
3516, (1999). In our case, perhaps the most interesting and
relevant form of disorder is the one associated
with disorder in vortex positions.
\bibitem{deGennes89} P.~G. de~Gennes,
{\em Superconductivity of Metals and Alloys} (Addison-Wesley, Reading, MA, 1989).
\bibitem{senthil2} T. Senthil, J. B. Marston, and M. P. A. Fisher, cond-mat/9902062.
\bibitem{luttinger} J. M. Luttinger, Phys. Rev. {\bf 135}, A1505 (1964).
\bibitem{obraz} Yu. N. Obraztsov, Fiz. Tverd. tela {\bf 6}, 414 (1964)
[Sov. Phys.-Solid State {\bf 6}, 331 (1964)]; {\bf 7}, 455 (1965).
\bibitem{smst} L. Smr\v{c}ka and P. St\v{r}eda, J. Phys. C {\bf 10} 2153 (1977).
\bibitem{streda} H. Oji and P. Streda, \prb {\bf 31}, 7291 (1985).
\bibitem{mahan} G. D. Mahan, {\em Many-Particle Physics} (Plenum, New York, 1990).
\bibitem{kohmoto} M. Kohmoto, Annals of Physics {\bf 160}, 343 (1985).
\bibitem{girvin} M. Jonson and S. M. Girvin, \prb {\bf 29}, 1939 (1984).
\bibitem{noncrossing} J. von Neumann and
E. Wigner, {\it Physik. Z.}, {\bf 30} 467 (1927).
\bibitem{mreedbsimon} M. Reed and B. Simon, {\em Fourier Analysis and
Self-Adjointess} (Academic, New York, 1975).
\bibitem{richtmyer} R. Richtmyer, {\em Principles of Advanced Mathematical Physics,
Vol.I,} (Springer-Verlag, New York, 1978).
\bibitem{sousa} P. de Sousa Gerbert, \prd {\bf 40}, 1346 (1989).
\bibitem{sousajackiw} P. de Sousa Gerbert and R. Jackiw, Commun. Math. Phys.{\bf 124}, 229 (1989).
\bibitem{sitenko1} Y. A. Sitenko and D. G. Rakityanskii, Phys. Atom. Nucl.{\bf 60}, 1497 (1997).
\bibitem{tarrach} C. Manuel and R. Tarrash, Physics Letters {\bf B 301}, 72 (1993).
\bibitem{regularizations}
M. G. Alford, J. March-Russell, and
F. Wilczek, Nuclear Physics {\bf B 328}, 140 (1989).
C. R. Hagen, \prl {\bf 64}, 503 (1990).
E. G. Flekk\o y and J. M. Leinaas, Int. J. Mod. Phys. {\bf A6}, 5327 (1991).
V. D. Skarzhinsky and J. Audretsch, J. Phys. {\bf A30} 7603 (1997).
C. G. Beneventano, M. De Francia and E. M. Santangelo, Int. J. Mod. Phys. {\bf A14} 4749, (1999).
\bibitem{ft98} M. Franz and Z. Te{\v s}anovi{\' c}, \prl {\bf 80}, 4763 (1998).
\bibitem{dsz} S. Dukan and Z. Te{\v s}anovi{\' c}, \prb {\bf 49}, 13017 (1994).
Z. Te{\v s}anovi{\' c} and P. Sacramento, \prl {\bf 80}, 1521 (1998).
\bibitem{poople} C. Poople {\em et. al.}, {\em Superconductivity}
(Academic Press, San Diego, 1995).
\bibitem{ambegaokar} V. Ambegaokar and A. Griffin, \prb {\bf 137}, A1151 (1965).
\end{references}
\end{document}